\documentclass[aps,prd,nofootinbib,10pt,tightenlines,twocolumn,noti,floatfix,superscriptaddress,showkeys]{revtex4-2}

\usepackage[utf8]{inputenc}          
\usepackage{graphicx}         
\usepackage[usenames,dvipsnames]{xcolor}

\usepackage{orcidlink}
\usepackage{amsmath,amssymb,amsfonts,slashed} 
\usepackage{mathtools}          
\usepackage{mathrsfs}
\usepackage{txfonts}
\usepackage{bm}
\usepackage{stmaryrd}
\usepackage{tensor}
\usepackage[utf8]{inputenc}
\usepackage{comment}
\usepackage{epsfig}
\usepackage{epstopdf}
\usepackage{enumitem,booktabs}
\usepackage[stable]{footmisc}
\usepackage{cleveref}

\definecolor{mred}{RGB}{127,0,25}
\definecolor{mdgr}{RGB}{51,51,51}
\definecolor{mag}{RGB}{211, 54, 130}
\definecolor{verm}{RGB}{164, 25, 0}

\hypersetup{colorlinks=true,
            citecolor=NavyBlue,
            linkcolor=NavyBlue,
            urlcolor=NavyBlue}
\usepackage{siunitx}                                  
\DeclareSIUnit{\fm}{\femto\metre}                     

\newcommand{\lie}{\mathscr{L}}

\newcommand{\psym}{\mathbb{P}_{\rm Sym}}
\newcommand{\piter}{\mathbb{P}_{i{\rm Even}}}
\newcommand{\pglob}{\mathbb{P}_{g{\rm Even}}}

\newcommand{\Ceven}{\rmC_{\rm Even}}
\newcommand{\Citer}{\rmC_{ i{\rm Even}}}

\newcount\hh
\newcount\mm
\mm=\time
\hh=\time
\divide\hh by 60
\divide\mm by 60
\multiply\mm by 60
\mm=-\mm
\advance\mm by \time
\def\hhmm{\number\hh:\ifnum\mm<10{}0\fi\number\mm}

\def\be{\begin{equation}}
\def\ee{\end{equation}}

\newcommand{\rmd}{\mathrm{d}}
\newcommand{\rmC}{\mathrm{C}}
\newcommand{\rmD}{\mathrm{D}}
\newcommand{\rmDD}{\mathbb{D}}
\newcommand{\rmS}{\mathrm{S}}

\usepackage[normalem]{ulem}

\begin{document}



\title{Conservative and dissipative sectors in a nonlinear scalar model for the gravitational self-force problem}

\author{Francisco M. Blanco\orcidlink{0000-0002-7711-8395}}
\affiliation{Max Planck Institute for Gravitational Physics (Albert Einstein Institute), Am M\"{u}hlenberg 1, Potsdam, 14476, Germany}

\author{Éanna É. Flanagan\orcidlink{0000-0001-6818-3550}}
\affiliation{Department of Physics, Cornell University, Ithaca, NY 14853, USA.}

\author{Abraham I. Harte\orcidlink{0000-0002-5893-5680}}
\affiliation{Centre for Astrophysics and Relativity (CfAR),
	School of Mathematical Sciences, \\
	Dublin City University, Glasnevin, Dublin 9, Ireland}

\begin{abstract}

  When considering how self-interaction affects an object's motion, it can be convenient to decompose the self-force into conservative and dissipative pieces. As a toy model for understanding such decompositions of the gravitational self-force, we consider objects that do not affect the spacetime, but are instead coupled to a nonlinear scalar field. There is then a standard splitting of the first-order scalar self-force into conservative and dissipative components. Multiple criteria can be used to obtain this splitting, all of which imply the same result. However, the implications of these criteria generically differ at higher orders. Demanding that any reasonable conservative sector be Hamiltonian, we identify multiple possible definitions of the conservative second-order self-force. Motivations for these possibilities and their properties are discussed and relevant Hamiltonians are obtained. We assume the existence of a three-point function with certain properties that is a generalization of the Detweiler-Whiting two-point function. These results apply to the two-body problem but are restricted to unbound scattering trajectories, due to infrared divergences that arise for bound orbits.

\end{abstract}

\maketitle

\vspace{0.1cm}

\section{\label{sec:intro} Introduction}

In recent decades, the fields of astrophysics and general relativity have experienced a revolution, thanks to groundbreaking advancements in gravitational-wave detection and analysis. Key instruments, such as the Laser Interferometer Gravitational-Wave Observatory (LIGO) \cite{introLIGO1, introLIGO2, introLIGO3}, the North American Nanohertz Observatory for Gravitational Waves (NANOGrav) \cite{Agazie_2023,Agazie2_2023}, and the upcoming Laser Interferometer Space Antenna (LISA) \cite{Audley:2017drz} have played -- or are expected to play -- pivotal roles in our ability to observe and study gravitational waves. Binary systems, mutually orbiting pairs of massive objects such as black holes or neutron stars, have been the primary sources for the gravitational waves observed thus far. To harness the information encoded in these waves, a diverse array of mathematical and computational methods have been developed. These include the perturbative approaches of the post-Newtonian \cite{introPN, poissonwill, Levi:2018nxp, Porto:2016pyg} and the post-Minkowskian  \cite{Damour:2016gwp} approximations, as well as the non-perturbative simulations employed in numerical relativity \cite{Lehner:2014asa}. Our ability to probe the dynamics of binary systems in different regimes has also been expanded by the effective one-body formalism \cite{Damour:2012mv, Taracchini:2013rva}, and by techniques tailored for systems with small mass ratios \cite{introEMRI, pound}.

Although binary systems are associated with an infinite number of radiative degrees of freedom, there are regimes in which those degrees of freedom can be ``integrated out.'' Doing so and then taking a point-particle limit results in a finite-dimensional dynamical system that describes the motion of the two bodies.
However, these dynamical systems generically contain nonlocal-in-time interactions, commonly referred to as tail effects. Tail effects imply that the relevant equations of motion are integro-differential equations rather than ordinary differential equations. Nevertheless, if all nonlocalities are treated perturbatively, one can derive a perturbatively-equivalent dynamical system that is completely local. The dynamics of such a system is characterized, at each order, by a vector field on a finite-dimensional phase space. In the context of such local dynamical systems, one can define a splitting into two sectors to mean a decomposition of this vector field into a sum of two terms.

One often encounters, for example, splittings into what are called ``conservative'' and ``dissipative'' sectors. While conservative-dissipative splittings are well-understood at leading order, there is no generally accepted consensus for how they might be defined more generally. A variety of criteria can be used. For example:

\begin{enumerate}[label=\Roman*)]

    \item The process of integrating out a radiative field depends on a choice of time orientation for that field. Here a reversal of the time orientation is defined by replacing retarded Green's functions with advanced Green's functions and vice versa. Given this definition, one can define a conservative sector to be a piece of the dynamical vector field that is unaffected by time orientation reversals.

    \item One can demand that the conservative sector constitutes a Hamiltonian dynamical system, so that there exists a Hamiltonian and a symplectic form on the phase space that reproduces the conservative dynamics. The conservative dynamics would then conserve at least the value of the Hamiltonian.

    \item For closed, bound orbits, one can define a conservative dynamics by demanding that there exist quantities -- energy, angular momentum, and so on -- that are conserved at least in an orbit-averaged context. The secular evolution of an orbit would then be driven just by the dissipative piece of the dynamics to leading adiabatic order.  

    \item  A closely related criterion is to impose that there exists a set of ``fluxes'' of energy, angular momentum etc. that can be computed from fields at future null infinity and at horizons, which reproduce the effect of the dissipative piece of the self-force in an orbit-averaged sense.

\end{enumerate}

In linear theories, the identification of conservative and dissipative sectors is unambiguous, and is typically based on the time-even criterion described in item I above. In a linear theory with a scalar field $\phi$, the retarded and the advanced fields $\phi^+$ and $\phi^-$ can be used to construct the conservative and the dissipative fields 
\begin{align}
        \phi^\rmC \equiv \frac{1}{2}( \phi^+ + \phi^- ),
        \qquad
        \phi^\rmD \equiv \frac{1}{2} ( \phi^+ - \phi^- ).
\end{align}
These fields are time-even and time-odd, respectively. In the context of the first-order self-force acting on a small body, conservative and dissipative self-forces can be defined in terms of similarly-constructed time-even and time-odd pieces of a certain regularized self-field. The conservative dynamics are then Hamiltonian, as shown for spinless point particles in Ref.\ \cite{Blanco:2022mgd}, and for spinning particles to leading order in spin in Ref.\ \cite{Blanco:2023jxf}. 

Identifying conservative and dissipative sectors is more subtle in nonlinear theories, as the implications of the four criteria outlined above (and others) do not necessarily coincide. Each of those criteria can also be consistent with more than one splitting of conservative and dissipative sectors. The time-even constraint can, for example, be applied in different ways in perturbation theory. At second order, different applications of that constraint differ by time-even second-order terms sourced by products of time-odd first-order terms. It has been an open question which of these prescriptions is preferred, if any, and if they admit Hamiltonian descriptions.

\subsection*{Strategy and overview}

The purpose of this paper is to explore possible definitions for conservative and dissipative sectors of the dynamics of small bodies that are subject to self-interaction.  
We focus on criteria I and II, and variants thereof, and do not consider criteria III and IV any further.

We are motivated by the dynamics of gravitationally-bound binary systems in the small mass-ratio regime, in which case the metric perturbation of the smaller object is treated as a perturbation to the background spacetime, expandable in powers of the mass ratio. The interaction of the small object with its own gravitational field results in a deviation from geodesic motion in the background. This deviation is captured by the gravitational self-force, which has first-order \cite{misata, quwa} and second-order contributions \cite{pound1, pound2, gralla, pound3, pound4, pound5, gralla}. As a toy model for such systems, we consider objects coupled to scalar fields with nonlinear self-interactions. We work to second order in the scalar charge of the body, which provides a model for the second-order gravitational self-force. 

The key questions we set out to address are: First, are there any definitions for the conservative second-order self-force that satisfy the Hamiltonian property II? Second, can one argue that there is a natural and unique definition for this component of the self-force? We will argue that natural\footnote{By natural, we mean that the definitions do not involve any structures on the phase space other than those that are already present.} decompositions exist for which the conservative sector is Hamiltonian, but that they are not unique.

There are multiple reasons for pursuing definitions that satisfy the Hamiltonian property II. Hamiltonian formulations provide a powerful tool set to study integrability, chaos, gauge-invariant quantities, and resonance effects \cite{Blanco:2022mgd}. They also offer a unifying language to
cross-validate results across different approximation schemes, as Hamiltonian formulations of conservative binary dynamics have been derived, in gravitational contexts, in both the post-Newtonian and the post-Minkowskian approximations; see Ref.\ \cite{hamiltonians} and references therein. Aspects of the motion that do not involve self-interaction can also be noted to be Hamiltonian. This is well-known for geodesics in a fixed background spacetime, and it remains true also when spin effects are included to leading order \cite{witzanyham, paulramond}.

Given a possible splitting of the dynamics into two sectors, it is not typically trivial to determine if the putative conservative sector admits a Hamiltonian description. This is because the vector field on phase space that determines the dynamics generically involves the nonlocal tail interactions. However, it was shown recently by one of us that a large class of finite-dimensional dynamical systems with perturbative, nonlocal-in-time interactions are Hamiltonian to all orders in the perturbative expansion  \cite{blanco2024localhamiltoniandynamicsnonlocal}. This result provides a general framework for deriving Hamiltonian formulations. The main result we use from this framework is that in order for the dynamics to be Hamiltonian, the fields that determine the conservative self-force should be constructed from fully-symmetric $n$-point functions.

Before using this formalism, we need equations of motion that incorporate all second-order self-force effects.
There have been several derivations of the second-order gravitational self-force, some involving matched asymptotic expansions, and others using so-called puncture schemes \cite{pound1, pound2, gralla, pound3, pound4, pound5, gralla}. However, because of the complexity of these formulations, it has not been possible to combine the resulting prescriptions for the second order self-force with the Hamiltonian formalism of Ref.\ \cite{blanco2024localhamiltoniandynamicsnonlocal}.
In this paper, we instead employ a formulation of nonlinear self-force effects that has recently been developed \cite{harte2025nonlinearly,harte2025nonlinearly1}, which is an extension of the nonperturbative method developed by one of us \cite{Harte_2008, harteeffectivemetric, Harte_2015}. The method computes the self-force on finite bodies using a nonlocal field transformation that preserves the form of the equations of motion while removing short lengthscales. Using this method, it is relatively straightforward to take point-particle limits. The resulting equations of motion are formally identical to those of a test body under the influence of an effective source-free field.

This paper provides multiple definitions for the conservative piece of the second-order scalar self-force which yield Hamiltonian dynamics. It does so by combining the description of the dynamics provided by Refs.\ \cite{harte2025nonlinearly, harte2025nonlinearly1}, with the general result of Ref.\ \cite{blanco2024localhamiltoniandynamicsnonlocal} that certain kinds of theories with nonlocal-in-time interactions are Hamiltonian order by order. Sec.\ \ref{sec:scalarmodel} uses the self-force results of Refs.\ \cite{harte2025nonlinearly, harte2025nonlinearly1}
to derive the second-order point-particle self-force associated with a nonlinear scalar field. This derivation assumes the existence of a certain 3-point function which we do not explicitly construct. Sec.\ \ref{sec:cons1st} reviews the conservative-dissipative split for the first-order self-force. Sec. \ref{sec:cons2nd} then extends these considerations to second order, which amounts to selecting different $n$-point functions that are symmetric under exchange of arguments. Lastly, Sec.\ \ref{sec:Hamiltonian} derives Hamiltonians and symplectic forms for our conservative sectors. 

There are three appendices. Appendix \ref{App:notation} summarizes our notation.  Appendix \ref{sec:Hamproof} reviews the Hamiltonian formalism of Ref. \cite{blanco2024localhamiltoniandynamicsnonlocal}, which includes an explicit prescription for computing Hamiltonians and symplectic forms even in dynamical systems with nonlocal-in-time interactions. Lastly, Appendix \ref{sec:appendixA} uses an explicit model to argue that  the simplest conceivable definition for the conservative self-force is not viable due to an infrared divergence.


\section{Motion of a small charged body: second-order scalar self-force}
\label{sec:scalarmodel}

This section derives the second-order self-force for a point charge coupled to a class of nonlinear scalar fields. After choosing a matter model, as well as a particular class of scalar fields, we begin by briefly reviewing the general self-force results obtained in Refs.\ \cite{harte2025nonlinearly, harte2025nonlinearly1}. Those results are then applied to write the self-force in terms of certain 2- and 3-point functions, whose existence we assume. Finally, we take a formal point-particle limit.

\subsection{Modeling an extended charge}

Consider the motion of an extended body with stress-energy tensor $T^{\mu\nu}$ and scalar charge density $\rho$ in a background spacetime with the fixed metric $g_{\mu\nu}$. Both this stress-energy tensor and this charge density are assumed to be spatially compact. The body is also assumed to couple to a scalar field $\phi$ via the field equation
\be\label{eq:fieldeqret}
    \nabla^\mu \nabla_\mu \phi -\dot{V}(\phi)=\rho,
\ee
where $V(\phi)$ is a potential and $\dot{V} \equiv \rmd V/\rmd \phi$. Stress-energy conservation in this context implies that 
\begin{equation}\label{eq:bareTmunu}
    \nabla_\mu T^{\mu\nu}=-\rho \nabla^\nu \phi.
\end{equation}
Net forces and torques arise by appropriately integrating the right-hand side of this equation, which can be interpreted as a force density.

While there are many systems with these properties, one possibility is described by the action  
\be
\label{action00}
S = S_{\rm fluid} - \int \! \rmd v \left[ \frac{1}{2} \nabla^\mu
  \phi \nabla_\mu \phi + V(\phi) + \kappa n \phi \right],
\ee
where $\rmd v \equiv \rmd^4 x \sqrt{-g}$ is the usual volume element. The first term here is the action for a perfect fluid with 4-velocity $u^\mu$, particle number density $n$, energy density $\varepsilon$, and pressure $p$, which obeys the conservation law $\nabla_\mu( n u^\mu)=0$.  The $\kappa$ in the last term is a constant, which we assume to be positive in order to ensure an attractive interaction. 

Differentiating the fluid and the $\kappa n \phi$ terms in the action with respect to the metric produces the stress-energy tensor $T^{\mu\nu}$, while the charge density is given by $\rho = \kappa n$. Although this model is not necessary, it has the advantage that the effective scalar charge of the body (as discussed below) can be shown to be conserved in the regime we consider \cite{harte2025nonlinearly1}. This provides a closer analogy with the gravitational case, where the effective mass of a body is conserved in the monopole approximation. For other matter models, the net scalar charge is not necessarily conserved.

\subsection{Field mappings}

In a naive point-particle limit, both $\rho$ and $\phi$ diverge in the force density $-\rho \nabla_\nu \phi$. The limit of that force density does not exist even as a distribution, and it cannot be integrated to yield a net force. It is therefore not possible to apply a point-particle limit and then to compute a force. What must be done instead is to compute forces on a sequence of finite bodies with ever-smaller sizes. In this context, the majority of the self-force can be absorbed into redefinitions -- or renormalizations -- of the stress-energy tensor and the charge density. Such effects depend sensitively on details of an object's internal structure, and their lack of universality is reflected in the lack of a well-defined point-particle limit. However, self-forces that cannot be absorbed in this sense \textit{are} universal and do admit a well-defined point-particle limit. It is these latter effects that we focus on. The key idea of \cite{harte2025nonlinearly} is that it is possible to find transformations that remove, at the outset, all components of the field that do not contribute to the universal component of the self-force. What remains after such a transformation is well-behaved even in a point-particle limit. The transformation itself can then be interpreted as a regularization procedure.

Our transformations can be obtained in terms of an appropriate ``generating functional'' $W[\rho, \hat{\phi},g_{\mu\nu}]$. This functional produces an effective field $\hat{\phi}$, a ``dressed'' (or ``effective'' or ``renormalized'') charge density $\hat{\rho}$, and a dressed stress-energy tensor $\hat{T}^{\mu\nu}$, all of which are implicitly determined by \cite{harte2025nonlinearly}
\begin{subequations}
\label{xform}
\begin{align}
    \phi = \frac{1}{\sqrt{-g}}  \frac{ \delta W }{ \delta \rho }, \qquad \hat{\rho} = \frac{1}{\sqrt{-g}}  \frac{ \delta W }{ \delta \hat{\phi} },
    \\
    \hat{T}^{\mu\nu} = T^{\mu\nu} - \frac{2}{ \sqrt{-g} } \frac{\delta W}{\delta g_{\mu\nu}} + \rho \phi g^{\mu\nu}.
\end{align}
\end{subequations}
The primary result is that if $W$ is invariant with respect to diffeomorphisms, the ``bare'' statement of stress-energy conservation, Eq.\ \eqref{eq:bareTmunu}, is mapped into its ``dressed counterpart''
\begin{equation}
    \nabla_\mu\hat{T}^{\mu\nu}=-\hat{\rho}\nabla^\nu \hat{\phi}.
    \label{divThat}
\end{equation}
An effective body with stress-energy tensor $\hat{T}^{\mu\nu}$ thus moves, at least instantaneously, as though it were a body with charge density $\hat{\rho}$ immersed in the effective external field $\hat{\phi}$. 

This result is useful, in part, because we can find transformations in which the effective force density $-\hat{\rho}\nabla_\nu \hat{\phi}$ has a well-behaved point-particle limit, unlike its bare counterpart $-\rho \nabla_\nu \phi$. Essentially, we need only that $\nabla_\nu \hat{\phi}$ remain well-behaved in that limit. In most scenarios, this limiting behavior can be guaranteed by choosing $W$ such that $\hat{\phi}$ satisfies the source-free field equation
\begin{equation}
\nabla^\mu \nabla_\mu \hat{\phi} - \dot{V}(\hat{\phi}) = 0,
    \label{vacPhi}
\end{equation}
at least in a neighborhood of the body of interest. In addition to this requirement, certain quasilocality conditions must also be imposed on our transformation in order to ensure that i) the effective body is not much larger than the physical one, and ii) that the effective body does not depend on anything in the distant past or future.

\subsection{Second-order expansions}
\label{sec:2ndOrder}

We now  use this formalism to compute the scalar self-force through second order, by which we mean the force that is at most quadratic in the charge density. The first step is to expand the generating functional $W[\rho, \hat{\phi},g_{\mu\nu}]$ in powers of $\rho$, which results in
\begin{align}
    W &= \int \rmd v \rho \hat{\phi} + \frac{1}{2} \int \! \rmd v \rmd v' \rho \rho' G_2^\rmS(x, x'; \hat{\phi}] 
    \nonumber
    \\
    & {} + \frac{1}{3} \! \int \rmd v \rmd v' \rmd v'' \rho \rho' \rho'' G_3^\rmS(x,x',x''; \hat{\phi}] + O(\rho^4).
    \label{Wexpand}
\end{align}
Here, the 2- and 3-point functions $G_2^\rmS$ and $G_3^\rmS$ are proportional to the second and the third functional derivatives of $W$ with respect to $\rho$. They are fully-symmetric and diffeomorphism-covariant, and they are also functionals\footnote{These $n$-point functions are also functionals of the metric, although we suppress that dependence for brevity. Later in the paper, we evaluate $G_2^\rmS$ and $G_3^\rmS$ on a background field $\bar{\phi}$ rather than on the effective field $\hat{\phi}$. After doing so, we suppress the field dependence as well.} of $\hat{\phi}$. The ``S'' superscripts are used here to suggest that $G_2^\rmS$ and $G_3^\rmS$ are ``singular'' $n$-point functions in the sense of the Detweiler-Whiting decomposition\footnote{All $n$-point functions in this paper are singular in the ordinary mathematical sense. However, what matters in practice is what happens when an $n$-point function is integrated against an appropriate charge density. If $G_2^\rmS$ and $G_3^\rmS$ were integrated against a sequence of charge densities whose limit describes a point particle, the results would not be finite. This is the sense in which we describe $G_2^\rmS$ and $G_3^\rmS$ as ``singular.'' They are not, however, the only $n$-point functions we consider with this property.} \cite{Detweiler:2002mi, introEMRI}. 

Substituting the expansion \eqref{Wexpand} for $W$ into Eq.\ \eqref{xform} shows that the effective field in which bodies appear to move is given by
\begin{align}
    \hat{\phi}(x) & = \phi - \int \! \rmd v' \rho' G_2^\rmS(x,x'; \hat{\phi}]
    \nonumber
    \\
    & {} - \int \! \rmd v' \rmd v'' \rho' \rho'' G_3^\rmS (x,x',x''; \hat{\phi}] + O(\rho^3).
    \label{phiHatexpand}
\end{align}
Similarly, a body's dressed charge density is given by
\begin{align}
    \hat{\rho} (x) = \rho(x) + \frac{1}{2} \int \! \rmd v' \rmd v'' \rho' \rho'' G_{2,1}^\rmS (x',x'';x;\hat{\phi}] + O(\rho^3),
    \label{rhoHat}
\end{align}
where
\begin{equation}
    G_{2,1}^\rmS (x',x'';x; \hat{\phi}] \equiv \frac{1}{\sqrt{-g}} \frac{ \delta G_2^\rmS (x',x'' ; \hat{\phi}] }{ \delta \hat{\phi}(x) }.
    \label{G21}
\end{equation}
Transformations generated by any symmetric and covariant $G_2^\rmS$ and $G_3^\rmS$ preserve the form of the laws of motion, in the sense that there is an effective stress-energy tensor that satisfies Eq.\ \eqref{divThat}. However, more restrictions are needed in order for these transformations to be useful.

As discussed above, it is, e.g., useful to demand that $\hat{\phi}$ be a solution to the source-free field equation \eqref{vacPhi}. Introducing
\begin{equation}
    \rmDD_{\hat{\phi}} \equiv \nabla^\mu \nabla_\mu - \ddot{V}(\hat{\phi}),
    \label{linDdef}
\end{equation}
which is the field operator linearized about $\phi = \hat{\phi}$, the effective field satisfies the source-free constraint when the $n$-point functions appearing in Eq.\ \eqref{Wexpand} are chosen such that
\begin{subequations}
\label{GhatEqs}
\begin{gather}
    \rmDD_{\hat{\phi}} G_2^\rmS (x,x'; \hat{\phi}] = \delta(x,x'),
    \\
    \rmDD_{\hat{\phi}} G_3^\rmS (x,x',x''; \hat{\phi}] = \frac{1}{2} \dddot{V}(\hat{\phi}) G_2^\rmS (x,x'; \hat{\phi}] G_2^\rmS (x,x''; \hat{\phi}].
\end{gather}
\end{subequations}
Besides being fully symmetric, satisfying these equations, and being covariantly constructed, our 2- and 3-point functions must also have the properties that \cite{harte2025nonlinearly, harte2025nonlinearly1}:
\begin{enumerate}
    \item $G^\rmS_2 (x,x'; \hat{\phi}]$  vanishes whenever $x$ and $x'$ are timelike-separated.
    \item $G^\rmS_{2,1} (x,x';x''; \hat{\phi}]$ vanishes when $x''$ lies outside of the geodesic connecting $x$ and $x'$. 
    \item $G^\rmS_3 (x,x',x''; \hat{\phi}]$ vanishes wherever two or more of the three possible pairs formed from $x$, $x'$, and $x''$ are timelike-separated.
\end{enumerate}
These conditions guarantee that the effective description of the system remains quasilocal. The effective variables describing the body do not depend on degrees of freedom distant from the body, nor on degrees of freedom in the distant past or future. The first condition identifies $G_2^\rmS$ as the Detweiler-Whiting ``singular'' 2-point function \cite{Detweiler:2002mi, introEMRI} appropriate to this theory. The second condition can be deduced from the Hadamard construction of that function. The third condition can be viewed as generalizing the Detweiler-Whiting prescription to higher order. In what follows, we assume the existence of 2- and 3-point functions with these properties. We do not, however, attempt to construct them. 

The $n$-point functions $G_2^\rmS$ and $G_3^\rmS$ provide a recipe for extracting that piece of the physical field $\phi$ that enters into a body's equations of motion. However, the physical field itself must be found separately. Choosing $\bar{\phi}$ to be a source-free background field, we expand this in powers of $\rho$ by letting 
\begin{equation}
    \phi = \bar{\phi} + \phi_{(1)} + \phi_{(2)} + O(\rho^3).
\end{equation}
The first-order and the second-order fields must then be solutions to 
\begin{equation}\label{eq:retfieldexpansion}
    \rmDD_{\bar{\phi}} \phi_{(1)} = \rho , \qquad \rmDD_{\bar{\phi}} \phi_{(2)} = \frac{1}{2} \dddot{V}(\bar{\phi}) \phi_{(1)}^2,
\end{equation}
where $\rmDD_{\bar{\phi}} = \nabla^\mu \nabla_\mu - \ddot{V}(\bar{\phi})$ is the field operator linearized around $\phi = \bar{\phi}$. We primarily focus on the retarded solution $\phi^+$, which can be constructed in terms of the retarded Green's function $G_2^+$ that satisfies
\begin{equation}\label{eq:ret2pointfieldeq}
    \rmDD_{\bar{\phi}} G_2^+(x,x'; \bar{\phi}] = \delta(x,x').
\end{equation} 
This Green's function depends on the background $\bar{\phi}$. It determines both $\phi_{(1)}$ and $\phi_{(2)}$, where the latter is constructed from the 3-point function
\begin{align}
    G_3^+(x,x',x'' ; \bar{\phi}] \equiv \frac{1}{2} \int \! \rmd v''' \dddot{V} ( \bar{\phi}''' ) G_2^+(x,x''' ; \bar{\phi}] 
    \nonumber
    \\
    {} \times  G_2^+(x''',x'; \bar{\phi}] G_2^+(x''',x''; \bar{\phi}].
    \label{G3plus}
\end{align}
In terms of $G_2^+$ and $G_3^+$, the retarded field is given by
\begin{align}
    \phi^+(x) &= \bar{\phi}(x) + \int \! \rmd v' \rho' G_2^+(x,x' ; \bar{\phi}]
    \nonumber
    \\
    & {} + \int \! \rmd v' \rmd v'' \rho' \rho'' G_3^+ (x,x',x'' ; \bar{\phi}] + O(\rho^3)
    \label{phiPlus}
\end{align}
through second order in $\rho$. This can be combined with \eqref{phiHatexpand} to find the associated effective field that appears in the laws of motion, which we denote by $\hat{\phi}^+$. 

There are two complications, however. First, we would like to express $\hat{\phi}^+$ in terms of the dressed charge density $\hat{\rho}$ rather than the bare charge density $\rho$. This can be accomplished by applying Eq.\ \eqref{rhoHat}. Second, we would like all $n$-point functions to be functionals of $\bar{\phi}$ rather than $\hat{\phi}$. Ignoring possible boundary terms and assuming that $\phi = \phi^+$, this can be accomplished by noting that
\begin{align}
    G_2^\rmS (x,x'; \hat{\phi}] = G_2^\rmS (x,x'; \bar{\phi}] + \int \! \rmd v'' \rmd v''' \rho'' \hat{G}_2^{+} (x''',x'' ; \bar{\phi}]
    \nonumber
    \\
    {} \times G_{2,1}^\rmS (x,x'; x'''; \bar{\phi}]  ,
    \label{G2Spert}
\end{align}
where 
\be
\label{eq:retreg2point}
    \hat{G}_2^{+}(x,x'; \bar{\phi}] \equiv G_2^+(x,x'; \bar{\phi} ] -G^\rmS_2(x,x'; \bar{\phi}]
\ee
is a solution to the linearized source-free field equation $\rmDD_{\bar{\phi}} \hat{G}_2^{+} = 0$. \textit{For the remainder of this paper, all $n$-point functions are to be understood as having been evaluated at $\bar{\phi}$.} For brevity, we omit this dependence in our notation, writing, e.g., $G_2^+(x,x';\bar{\phi}]$ as $G_2^+(x,x')$. 

The effective field $\hat{\phi}^+$ that is associated with a retarded physical field $\phi^+$ can now be found by combining Eqs.\ \eqref{phiHatexpand}, \eqref{rhoHat}, \eqref{phiPlus}, \eqref{G2Spert} and \eqref{eq:retreg2point}. Doing so results in
\begin{eqnarray}
\label{eq:hatphiexprextended}
    \hat{\phi}^+(x) &=& \bar{\phi}(x) + \int \! \rmd v' \hat{\rho}' \hat{G}_{2}^{+}(x,x')
    \nonumber \\
    &&
    +\int \! \rmd v' \rmd v'' \hat{\rho}'\hat{\rho}'' \hat{G}_3^{+}(x,x',x'') +
    O({\hat{\rho}}^3),
\end{eqnarray}
where the relevant 3-point function is given by
\begin{align}
    \hat{G}_3^{+}(x,x',x'') \equiv G_3^+ (x,x',x'') -G^\rmS_3(x,x',x'') 
    \nonumber
    \\
    {} - \int \! \rmd v''' \bigg[G^\rmS_{2,1}(x,x';x''') \hat{G}_2^{+}(x''',x'')
    \nonumber
    \\
    {} + \frac{1}{2} G^\rmS_{2,1}(x',x'';x''') \hat{G}_2^{+}(x,x''') \bigg].
\label{eq:retreg3point}
\end{align}
The effective field $\hat{\phi}^-$ that is associated with the advanced field $\phi^-$ would be given by the same expression, but with every retarded Green's function $G_2^+$ replaced with an advanced Green's function $G_2^-$. Recalling that the $G_2^\rmS$ and the $G_3^\rmS$ appearing here are evaluated on the background $\bar{\phi}$, they satisfy the field equations \eqref{GhatEqs} with the replacement $\hat{\phi} \to \bar{\phi}$. More explicitly, the singular $n$-point functions used in the remainder of this work satisfy
\begin{subequations}\label{eq:singularnpointeqs}
\begin{eqnarray}    
         \rmDD_{\bar{\phi}} G^\rmS_2(x,x')&=& \delta (x,x'), \\
         \rmDD_{\bar{\phi}} G^\rmS_{2,1}(x,x';x'') &=& \dddot{V}(\bar{\phi}) G^\rmS_2 (x,x') \delta(x, x'' ), \\
         \rmDD_{\bar{\phi}} G^\rmS_3(x,x',x'') &=& \frac{1}{2} \dddot{V}(\bar{\phi}) G^\rmS_2 (x,x') G^\rmS_2(x,x'').
\end{eqnarray}
\end{subequations}
The notation used for the various 2- and 3-point functions defined in this section is summarized in Appendix \ref{App:notation}.

Because the effective field ${\hat \phi}^+$ is source-free, it has a finite point-particle limit in most cases of physical interest. The hats on the $\hat{G}_2^{+}$ and $\hat{G}_3^{+}$ that determine $\hat{\phi}^+$ denote that they are in this sense ``more regular'' than $G_2^+$ and $G_3^+$. These $n$-point functions are, however, still singular; regularity arises only after integration against $\hat{\rho}$. To be more precise, we expect that a point-particle limit constructed by considering suitable 1-parameter families of extended bodies will produce a regular effective field. This requirement is not the same as directly asking for the formalism to make sense for a literal point particle, with no limiting process involved. Indeed, solutions to nonlinear field equations with point-particle sources -- such as the retarded field with such a source -- generically do not exist, at least within the ordinary theory of distributions \cite{Geroch1987}.

This limitation does not necessarily preclude an effective field from existing for a point particle source (which has not been obtained from a limiting procedure), and although we have not verified it, we assume below that it is meaningful to simply replace the $\hat{\rho}$ in \eqref{eq:hatphiexprextended} with a $\delta$-function concentrated on a worldline parameterized by $z_\tau$.  Formally doing so while assuming that $\tau$ represents a proper time produces the effective field
\begin{eqnarray}
\label{eq:hatphiexpr}
    \hat{\phi}^+(x)&=&\bar{\phi}(x) + \hat{q}\int \! \rmd \tau' \hat{G}_{2}^{+}(x,z_{\tau'})
    \nonumber \\
    &&
    {} + \hat{q}^2\int \! \rmd \tau' \rmd \tau'' \hat{G}_3^{+}(x,z_{\tau'},z_{\tau''})+
    O(\hat{q}^3),
    \label{phiHatPlus}
\end{eqnarray}
where $\hat{q}$ denotes the net effective charge. We assume that this charge is constant. As noted above, and as shown in \cite{harte2025nonlinearly1}, the constancy of $\hat{q}$ can be guaranteed by, e.g., adopting the microscopic matter model \eqref{action00}. Regardless, this $\hat{\phi}^+$ represents the point-particle effective field associated with a retarded physical field.  It takes into account only the monopole moment of the effective charge distribution. 

Similarly neglecting all multipole moments of the stress-energy tensor except for the monopole, the equation of motion is given by 
\be\label{eq:scalareomeff}
    \frac{\rmD}{\rmd\tau}\big(\hat{m} u^\mu\big)= -\hat{q}\nabla^\mu \hat{\phi}^+ \equiv f^\mu,
\ee
where $f^\mu$ denotes the force, $\hat{m}$ the dressed mass and $u^\mu= \rmd z^\mu_\tau/ \rmd \tau$ the 4-velocity. While $\hat{m}$ can vary, the equation of motion implies that $\hat{m} - \hat{q} \hat{\phi}^+$ is conserved. The effective field $\hat{\phi}^+$ that appears here is a functional of the body's worldline and admits the expansion 
\be
    {\hat \phi}^+ = {\bar \phi} + {\hat \phi}^+_{(1)} + {\hat \phi}^+_{(2)}+O(\hat{q}^3)
\ee
in powers of ${\hat q}$. The contribution of the background field $\bar{\phi}$ to $f^\mu$ is interpreted as the external force, the contribution of the first-order field
\begin{equation}
\label{fof}
    \hat{\phi}^+_{(1)}(x) = \hat{q}\int \! \rmd\tau' \hat{G}_{2}^{+}(x,z_{\tau'}) 
\end{equation}
is interpreted as the first-order self-force and the contribution of the second-order field 
\begin{equation}
    \hat{\phi}^+_{(2)}(x) = \hat{q}^2\int \! \rmd\tau' \rmd\tau'' \hat{G}_3^{+}(x,z_{\tau'}, z_{\tau''})
    \label{phiHat2}
\end{equation}
is interpreted as the second-order self-force.


\section{Conservative and dissipative sectors of the dynamics: First Order}\label{sec:cons1st}

We now review how to define conservative and dissipative sectors of a body's dynamics to first order in its charge. This requires definitions for the conservative and the dissipative pieces of the first-order contribution to the force $f^\mu$ that appears in Eq.\ \eqref{eq:scalareomeff}. More precisely, it requires that we define conservative and dissipative components of the first-order effective field $\hat{\phi}^+_{(1)}$ given by Eq.\ \eqref{fof}. That field depends on the 2-point function $\hat{G}_2^{+}(x,x')$, which can itself be split into conservative and dissipative components. 

First note that it is standard to split the retarded Green's function $G_2^+(x,x')$ into conservative and dissipative pieces via
\be
    G^+_2 = G_2^\rmC (x,x') + G_2^\rmD (x,x') ,
    \label{decompos1}
\ee
where 
\begin{subequations}
\label{eq:CD2pointa}
\begin{eqnarray}
    \label{eq:CD2pointCa}G_2^{\rmC}&\equiv&\frac{1}{2}\big[G_2^{+}(x,x')+G_2^{-}(x,x')],\\
    \label{eq:CD2pointDa}G_2^{\rmD}&\equiv&\frac{1}{2}\big[G_2^{+}(x,x')-G_2^{-}(x,x')].
\end{eqnarray}
\end{subequations}
The retarded and the advanced Green's functions satisfy the reciprocity relation \cite{introEMRI}
\begin{equation}
    G^+_2(x,x') = G^-_2 (x',x),
\end{equation}
so $G_2^\rmC$ is symmetric in its arguments while $G_2^\rmD$ is antisymmetric. Similarly, the effective retarded 2-point function $\hat{G}_2^+$ that is defined by Eq.\ (\ref{eq:retreg2point}) can be split into symmetric and antisymmetric pieces. Recalling that $G_2^\rmS$ is itself symmetric, the antisymmetric component is just\footnote{One way to understand the fact that the dissipative 2-point function is unmodified here is to note that it already satisfies $\rmDD_{\bar{\phi}} G^\rmD_2 = 0$; no subtractions are needed to make it a homogeneous solution.} $G_2^\rmD$, and that split results in 
\be
    \hat{G}^{+}_2 = \hat{G}_2^{\rmC}(x,x') + G_2^{\rmD} (x,x'),
\label{decompos2}
\ee
where
\begin{equation}
\label{eq:CD2point}
    \hat{G}_2^{\rmC} \equiv G_2^\rmC(x,x') - G_2^\rmS(x,x').
\end{equation}
The conservative piece of the first-order effective field ${\hat \phi}^+_{(1)}$ can then be defined as the field that is generated by integrating $\hat{G}_2^{\rmC}$ against the charge density:
\begin{equation}
    \hat{\phi}_{(1)}^\rmC (x) \equiv \hat{q} \int \! \rmd \tau' \hat{G}^\rmC_2 (x,z_{\tau'}).
    \label{phiCHat1}
\end{equation}
The conservative sector of the retarded first-order self-force is therefore
\be
  \label{eq:def2pointcons}
  f_{(1)\mu}^{\rmC} (\tau) \equiv -\hat{q}\nabla_\mu \hat{\phi}^{\rmC}_{(1)} (z_\tau)
    = -\hat{q}^2 \nabla_\mu\int \! \rmd\tau' \hat{G}_2^{\rmC} (z_\tau,z_{\tau'}).
\ee
Note that the same result would have been obtained had we begun with the advanced field rather than the retarded field.

A number of prescriptions can be used to arrive at this definition. Some examples are as follows:
\begin{enumerate}
    \item Select the piece of $\hat{G}^{+}_2$ which is obtained by symmetrizing under exchange of arguments. 
    \item Select the piece of $\hat{G}^{+}_2$ which is invariant under time reversal, meaning that it remains the same when retarded and advanced Green's functions are exchanged.
    \item Instead of finding an effective field for the retarded field and then isolating a particular component of that field, start by solving the field equations with time-symmetric boundary conditions and then find the effective field associated with that solution.
\end{enumerate}
These prescriptions all give the same result at first order. However, we show in the following section that analogous prescriptions do not agree at second order.

\section{Conservative and Dissipative Sectors of the Dynamics: Second Order}\label{sec:cons2nd}

This section explores how to define a conservative sector of the dynamics to second order in a body's scalar charge ${\hat q}$. As mentioned in the previous section, there are multiple reasonable prescriptions for the conservative dynamics at first order, and we now explain how these do or do not extend to second order. Throughout, we demand that any reasonable conservative sector admit a Hamiltonian description. As shown in  Appendix~\ref{sec:Hamproof}, this requires that any relevant $n$-point functions be fully symmetric under exchange of their arguments. However,  this restriction still allows more than one natural definition for the conservative second-order self-force.

\subsection{Projecting $n$-point functions}
\label{sec:projection}

The net force $f^\mu$ is linear in the effective field $\hat{\phi}^+$. All contributions to this field that affect the self-force also depend on certain $n$-point functions that are integrated against a body's dressed charge density ${\hat \rho}$, which we assume to be concentrated on the worldline parameterized by $z_\tau$. A conservative second-order self-force can thus be defined as the force associated with a field that has the form
\begin{equation}
  \label{eq:hatphi2C}
    \hat{\phi}_{(2)}^{\rmC} (x) \equiv \hat{q}^2 \int \! \rmd \tau' \rmd \tau'' \hat{G}_3^{\rmC}(x,z_{\tau'}, z_{\tau''}), 
\end{equation}
where $\hat{G}_3^{\rmC}$ is to be determined. Recalling Eq.\ \eqref{phiHat2}, the full second-order self-force depends on a field with this form, but with $\hat{G}_3^{\rmC}$ replaced by the $\hat{G}_3^{+}$ given by Eq.\ \eqref{eq:retreg3point}. This latter 3-point function does not, however, generate a conservative sector in any reasonable sense. It is also not symmetric in its arguments, so the associated dynamics are not Hamiltonian. Certain modifications to $\hat{G}_3^{+}$ must be performed in order to obtain a reasonable conservative sector.

These modifications can be explored by defining certain projection operators. Given Eqs.\ \eqref{G3plus}, \eqref{eq:retreg2point} and \eqref{eq:retreg3point}, the 3-point function $\hat{G}_3^{+}$ may be viewed as depending on the retarded Green's function $G_2^+$, as well as on $G_2^\rmS$ and $G_3^\rmS$. The retarded Green's function in turn depends on the physical boundary conditions. If we had used advanced boundary conditions instead, $G_2^+$ would be replaced by $G_2^-$ while $G_2^\rmS$ and $G_3^\rmS$ would remain as is. With this in mind, we consider objects with the form
\begin{equation}
    g_n(x_1, \ldots, x_n; G_2^+ , G_2^-] .
    \label{gn}
\end{equation}
These are $n$-point functions on spacetime that depend functionally on $G_2^+$ and $G_2^-$. Our primary examples are $\hat{G}_2^{+}$ for $g_2$ and $\hat{G}_3^{+}$ for $g_3$, although other possibilities can arise as well. Note that although these examples depend on, e.g., $G_2^\rmS$ and $G_3^\rmS$ as well as on $G_2^+$ and $G_2^-$, the former two $n$-point functions here are to be viewed as independent of the latter two.

We now introduce three different projection operators that act on the $n$-point functionals $g_n(x_1,\ldots, x_n; G_2^+, G_2^-]$: 
\begin{itemize}
    \item Define the symmetrization operator $\psym$, which symmetrizes $g_n$ over its $n$ arguments. Letting $s$ denote a permutation of the first $n$ integers $\{1 , \ldots , n\}$,
    \begin{equation}\label{eq:defprojsym}
        \psym \, g_n \equiv \frac{1}{n!} \sum_{s} g_n (x_{s(1)}, \ldots x_{s(n)}; G_2^+, G_2^-].
    \end{equation}

    \item Define the ``global time-even projection'' operator $\pglob$ such that 
    \begin{align}\label{eq:defprojeven}
        \pglob \, g_n \equiv \frac{1}{2} \Big\{ g_n (x_1, \ldots, x_n; G_2^+, G_2^-] 
        \nonumber
        \\
        ~ + g_n (x_1, \ldots, x_n; G_2^-, G_2^+] \Big\}.
    \end{align}
    This averages the original $n$-point function with one that has the opposite time orientation. 
    
    \item Define the ``iterated time-even projection'' operator $\piter$ such that
    \begin{equation}\label{eq:defprojitereven}
        \piter \, g_n \equiv g_n(x_1, \ldots, x_n; G^\rmC_2 , G^\rmC_2 ] .
    \end{equation}
    This replaces all occurrences of $G_2^+$ and $G_2^-$ with $G_2^\rmC = (G_2^+ + G_2^-)/2$, which is the average of both functional arguments of $g_n$. It enforces time-even dynamics at each order in perturbation theory, hence our appellation ``iterated time-even projection.''

\end{itemize}
These operators are all projections in the sense that applying any one of them twice is the same as applying it once. All three operators also commute with one another. Furthermore we have the identity
\begin{eqnarray}
    \pglob \piter = \piter.
    \label{projidentities}
\end{eqnarray}
If an application of $\pglob$ leaves an $n$-point function unchanged, it might be described as ``time-even.'' Both $\pglob$ and $\piter$ thus produce time-even $n$-point functions. However, these operators can produce distinct time-even results. The ``global'' projection $\pglob$ acts to make an entire expression time-even, even if it is constructed using building blocks -- the advanced and the retarded Green's functions -- that are not themselves time-even. By contrast, the ``iterated'' time-even projection $\piter$ works by making the building blocks themselves time-even.

All three projectors applied to the 2-point function $\hat{G}_2^{+}$ have the same effect, meaning that
\begin{equation}
    \psym \hat{G}_2^{+} = \piter \hat{G}_2^{+} = \pglob \hat{G}_2^{+} = \hat{G}_2^{\rmC}.
\end{equation}
A similar comment also applies to $\hat{G}_2^-$. All three operations can thus be applied to the first-order 2-point function in order to produce the same first-order conservative self-force.

Things are more interesting at second order. In order to understand this, it is first convenient to abuse the notation slightly by applying our operators to fields rather than $n$-point functions. If a field is computed using integrals of different $n$-point functions, a projector applied to the field is defined to mean that those same integrals are to be evaluated, but with the projector applied to all relevant $n$-point functions. For example, recalling Eq.\ \eqref{phiHatPlus},
\begin{align}
    \psym \hat{\phi}^+(x)& = \bar{\phi}(x) + \hat{q}\int \! \rmd \tau' \psym \hat{G}_{2}^{+}(x,z_{\tau'})
    \nonumber \\
    & {} + \hat{q}^2\int \! \rmd \tau' \rmd \tau'' \psym \hat{G}_3^{+}(x,z_{\tau'},z_{\tau''})+
    O(\hat{q}^3).
\end{align}

More generally, we can use various combinations of the projection operators to define conservative effective fields ${\hat \phi}^\rmC$ with the forms
\be
    {\hat \phi}^\rmC = \psym^a \, \piter^b \, \pglob^c \, {\hat
  \phi}^+.
  \label{eq:conseff1}
\ee
The fact that all operators here are commuting projections means that there is no loss of generality in restricting each of the exponents $a$, $b$ and $c$ to be equal to $0$ or to $1$, and in restricting to the order of projections indicated.
For each possibility, we define the associated dynamics by
\be\label{eq:conservativeeom}
\frac{\rmD}{\rmd\tau}\big(\hat{m}u^\mu\big)=-\hat{q}\nabla^\mu \hat{\phi}^\rmC.
\ee
This can be interpreted as a conservative sector for some values of $a$, $b$, and $c$. Expanding the conservative effective field $\hat{\phi}^\rmC$ to second order, its first-order contribution is given by Eq.\ \eqref{phiCHat1}, while its second-order contribution is given by replacing the 3-point function in Eq.\ (\ref{phiHat2}) with the projection $\psym^a \piter^b \pglob^c \hat{G}^+_3$.

Given that the exponents $a$, $b$ and $c$ can each be equal to 0 or to 1, Eq.\ \eqref{eq:conseff1} describes at most eight possible conservative effective fields. Demanding that the dynamics be Hamiltonian requires that any 3-point function be symmetric, so there is no loss of generality in setting $a=1$. This leaves us with four possibilities. Two of these possibilities are equivalent due to Eq.\ \eqref{projidentities}, and what remains are only
\begin{equation}
    \psym, \qquad \psym \piter, \qquad \psym \pglob.
\end{equation}
These projections are explored in the subsections below and are summarized in Table \ref{tab:Prescriptions}.

Our discussion thus far has focused only on the conservative sector. Any such prescription implicitly fixes the dissipative sector as that portion of the full dynamics that remains after subtracting the conservative contribution. At first order, the dissipative two-point function is antisymmetric under exchange of its arguments and is odd under time reversal. At higher orders, however, the dissipative $n$-point functions are not necessarily odd under time reversal, nor are they fully antisymmetric under exchange of arguments.

\subsection{Time-even conservative sector}

We begin our exploration of possible conservative sectors by considering the use of the global time-even projector
$\pglob$. If this is applied on its own to the 3-point function $\hat{G}_3^+$ given by Eq.\ \eqref{eq:retreg3point}, the result is not symmetric and cannot be associated with a Hamiltonian description. We must therefore also apply $\psym$, which results in the time-even conservative effective field 
\be
   {\hat \phi}^{\Ceven} \equiv \psym \pglob  {\hat \phi}^+ = \psym \pglob  {\hat \phi}^-.
   \label{phiEven}
\ee

\noindent
Here, we have labeled the \textit{time-even conservative sector}
\footnote{Both $\pglob$ and $\piter$ produce time-even fields. However, we argue in Sec.\ \ref{sec:noIEven} below that the field associated with $\piter$ is infrared-divergent. We therefore drop the adjective global here and refer to the field associated with $\pglob$ as \textit{the} time-even field.} 
by ``$\Ceven$.'' Using Eqs.\ \eqref{G3plus}, (\ref{eq:retreg3point}), (\ref{decompos1}) and (\ref{decompos2}), the corresponding 3-point function 
\begin{equation}
\label{eq:Ce}
\hat{G}^{\Ceven}_3 \equiv \pglob \psym  \hat{G}^{+}_3
\end{equation}
is explicitly given by
\begin{widetext}
\begin{align}
  \label{eq:Ceexp}
      \hat{G}^{\Ceven}_3(x,x',x'') = \frac{1}{2} \int \rmd v''' \Bigg\{ \dddot{V}(\bar{\phi}''') \bigg[ G_2^{\rmC}(x,x''') G_2^{\rmC}(x', x''') G_2^{\rmC}(x'', x''')
    - \frac{1}{3} \bigg( G^\rmD_2(x,x''') G^\rmD_2(x', x''') G^\rmC_2(x'', x''') 
    \nonumber
    \\
    {} +  G^\rmD_2(x,x''') G^\rmC_2(x', x''') G^\rmD_2(x'', x''')   + G^\rmC_2(x,x''') G^\rmD_2(x', x''') G^\rmD_2(x'', x''')  \bigg) \bigg] - G^\rmS_{2,1}(x,x';x''') \hat{G}_2^{\rmC}(x'', x''')
    \nonumber
    \\
    {} - G^\rmS_{2,1}(x,x'';x''') \hat{G}_2^{\rmC}(x', x''')
    - G^\rmS_{2,1}(x',x'';x''') \hat{G}_2^{\rmC}(x,x''') \Bigg\} -G^\rmS_3(x,x',x'').
\end{align}
\end{widetext}
Note that this depends on the dissipative 2-point function $G_2^\rmD = \frac{1}{2} (G_2^+ - G_2^-)$, which is not itself time-even. However, all such objects arise in pairs, and those pairs \textit{are} time-even.

The singular $n$-point functions $G_2^\rmS$ and $G_3^\rmS$ allow for transformations between physical fields, such as $\phi^+$, and effective fields, such as $\hat{\phi}^+$. The latter satisfies the homogeneous field equation \eqref{vacPhi}, and is thus expected to be finite even in a point-particle limit. We now verify that our conservative projections are similarly well-behaved. First note that if the linearized wave operator $\mathbb{D}_{\bar{\phi}}$ is applied to the time-even conservative 3-point function given by Eq.\ \eqref{eq:Ceexp}, the field equations \eqref{eq:ret2pointfieldeq} and \eqref{eq:singularnpointeqs} imply that
\begin{align}
    \mathbb{D}_{\bar{\phi}}  \hat{G}^{\Ceven}_3 (x,x',x'') =\frac{1}{6} \dddot{V}(\bar{\phi}) \Big[ 3 \hat{G}^{\rmC}_2 (x,x') \hat{G}^{\rmC}_2 (x,x'')
     \nonumber
     \\
     {}- G^{\rmD}_2(x,x') G^{\rmD}_2(x,x'') \Big].
\end{align}
Integrating this against the charge density shows that the associated second-order conservative effective field satisfies
\begin{equation}\label{eq:sourceCeven}
    \mathbb{D}_{\bar{\phi}} \hat{\phi}_{(2)}^{\Ceven} = \frac{1}{6}  \dddot{V} (\bar{\phi} ) \left[ 3 \big( \hat{\phi}_{(1)}^{\rmC} \big)^2 -\big( \phi_{(1)}^{\rmD} \big)^2 \right],
\end{equation}
where
\begin{equation}
    \phi^\rmD_{(1)} (x) \equiv \hat{q} \int \! \rmd \tau' G_2^\rmD (x,z_{\tau'}).
    \label{phiD}
\end{equation}
Both $\hat{\phi}^\rmC_{(1)}$ and $\phi^\rmD_{(1)}$ are solutions to homogeneous linearized field equations, and are therefore expected to be finite on the worldline. The source for $\hat{\phi}_{(2)}^{\Ceven}$ is therefore finite, suggesting that the field is itself well-behaved.

Combining both the first-order and the second-order fields, note that our conservative time-even effective field is not a solution to the homogeneous wave equation. Instead,
\begin{align}
    \nabla^\mu \nabla_\mu \hat{\phi}^{\Ceven} - \dot{V} \big( \hat{\phi}^{\Ceven} \big) = - \frac{1}{6} \dddot{V}(\bar{\phi}) \big( \phi_{(1)}^\rmD \big)^2 + O \big( \hat{q}^3 \big).
\end{align}
Although the analogous equation for $\hat{\phi}^+$ is source-free, the finiteness of the source on the right-hand side of this equation implies that its presence is entirely benign.

\subsection{Symmetrized conservative sectors}

Invariance under time reversal has traditionally been a requirement that has been imposed in order to define conservative sectors of dynamical systems. Perhaps  surprisingly, it is not necessary to satisfy this requirement in order to obtain a Hamiltonian dynamical system. We can define the \textit{symmetrized conservative sector} associated with the retarded field, denoted by ``$\rmC_+$,'' via
\be
   {\hat \phi}^{\rmC_+} \equiv \psym  {\hat \phi^+}.
   \label{phiSym}
\ee
The 3-point function associated with the symmetrized conservative sector is then
\begin{equation}
\label{eq:Cpm}
    \hat{G}^{\rmC_+}_3  \equiv \psym  \hat{G}^{+}_3 .
\end{equation} 
Another symmetrized conservative sector, denoted by ``$\rmC_-$,'' can be defined by starting with the advanced effective field $\hat{\phi}^-$ rather than the retarded one\footnote{More generally, it is possible to define a one-parameter family of conservative sectors by applying the projection operator $\psym$ to the 3-point functions $\alpha \hat{G}^{+}_3 + (1-\alpha) \hat{G}^{-}_3$, where $0 \le \alpha \le 1$. All of these sectors admit Hamiltonian descriptions.}:
\be
    \label{eq:Ce11}
    \hat{G}^{\rmC_-}_3 \equiv \psym  \hat{G}^{-}_3 .
\end{equation}
In both cases, the symmetry of the resulting 3-point functions is sufficient to ensure that the dynamics admit Hamiltonian descriptions. Explicitly, we can use Eqs.\ \eqref{G3plus}, (\ref{eq:retreg3point}), (\ref{decompos1}), (\ref{decompos2}) and (\ref{eq:Ceexp}) to write $\hat{G}^{\rmC_+}_3$ and $\hat{G}^{\rmC_-}_3$ as  $\hat{G}^{\Ceven}_3$ plus corrections:
\begin{widetext}
\begin{align}\label{eq:Cpmexp}
      \hat{G}^{\rmC_\pm}_3(x,x',x'') = \hat{G}^{\Ceven}_3(x,x',x'')
        \mp \frac{1}{6} \int \! \rmd v''' \bigg[ \dddot{V}(\bar{\phi}) \Big( G^\rmC_2(x,x''') G^\rmC_2(x',x''') G^\rmD_2(x'',x''') +  G^\rmD_2(x,x''') G^\rmC_2(x',x''') G^\rmC_2(x'',x''') 
       \nonumber
       \\
       {} + G^\rmC_2(x,x''') G^\rmD_2(x',x''')  G^\rmC_2(x'',x''')
        -3 G^\rmD_2(x,x''') G^\rmD_2(x',x''')G^\rmD_2(x'',x''') \Big)
        - G^\rmS_{2,1}(x,x';x''')G_2^{\rmD}(x'',x''')
        \nonumber
        \\        
        {} - G^\rmS_{2,1}(x,x'';x''') G_2^{\rmD}(x',x''')
        - G^\rmS_{2,1}(x',x'';x''')G_2^{\rmD}(x,x''') \bigg].
\end{align}
\end{widetext}
The terms here that do not appear in $\hat{G}_3^{\Ceven}$ include odd numbers of dissipative 2-point functions, which makes them sensitive to the choice of time orientation. Indeed, the signs of these terms depend on whether we project the retarded or the advanced solution. This dependence on time orientation is unusual, but not a sign of any inconsistency: the two corresponding Hamiltonians inherit preferred time orientations from the underlying retarded or advanced dynamics.

Like its time-even counterpart, the symmetrized conservative sector has a finite source, and is therefore expected to have a well-behaved point-particle limit. Use of Eq.\ (\ref{eq:singularnpointeqs}) shows that
\begin{equation}\label{eq:sourceCpm}
    \mathbb{D}_{\bar{\phi}
    } \hat{\phi}_{(2)}^{\rmC_\pm} = \frac{1}{6} \dddot{V}(\bar{\phi}) \Big[ 3 \big( \hat{\phi}_{(1)}^{\rmC} \big)^2 - \big(\phi_{(1)}^{\rmD} \big)^2
     \pm 2 \phi_{(1)}^{\rmD}\hat{\phi}_{(1)}^{\rmC} \Big].
\end{equation}
Both $\hat{\phi}^\rmC_{(1)}$ and $\phi^\rmD_{(1)}$ are homogeneous solutions to the linearized field equation, so the source here is finite. We therefore expect $\hat{\phi}_{(2)}^{\rmC_\pm}$ to be finite as well. The full symmetrized conservative effective fields can be shown to be solutions to the inhomogeneous field equation
\begin{align}
    \nabla^\mu \nabla_\mu  \hat{\phi}^{\rmC_\pm} - \dot{V} \big( \hat{\phi}^{\rmC_\pm} \big) = \frac{1}{6} \dddot{V} (\bar{\phi}) \phi_{(1)}^{\rmD} \big( \pm 2  \hat{\phi}_{(1)}^{\rmC} - \phi_{(1)}^{\rmD} \big) + O\big ( \hat{q}^3 \big) ,
\end{align}
which also has a finite source.

\subsection{No consistent conservative sector from the iterated time-even projector}
\label{sec:noIEven}

We have now discussed two of the three possible conservative sectors defined by projections whose forms are given by Eq.\ \eqref{eq:conseff1}. The last of these projections involves the iterated time-even projector $\piter$. Applying this to $\hat{G}^+_3$ does not result in a symmetric 3-point function, so we symmetrize the result and define the \textit{iterated time-even conservative sector} in terms of
\be
   {\hat \phi}^{\Citer} \equiv \psym \piter {\hat \phi}^+ = \psym \piter {\hat \phi}^-,
   \label{phiIeven}
\ee
with the corresponding 3-point function being given by
\begin{equation}
\label{eq:Cm}
\hat{G}^{\Citer}_3 \equiv \psym \piter \hat{G}^{+}_3 .
\end{equation}
This essentially corresponds to solving the field equations, at each order, with the conservative Green's function $G^\rmC_2$ rather than the retarded Green's function $G^+_2$. In this sense, it is conceptually the simplest possible proposal for a conservative sector. However, it does not yield a viable definition.

Temporarily leaving aside the reasons for this lack of viability, the same types of calculations as those that led to Eqs.\ \eqref{eq:Ceexp} and \eqref{eq:Cpmexp} can be used to show that the 3-point function here is at least formally given by
\begin{widetext}
\begin{align}
    \hat{G}_3^{\Citer}(x,x',x'') = \frac{1}{2} \int \! \rmd v''' \Big[ \dddot{V} (\bar{\phi}''') G_2^{\rmC}(x,x''')G_2^{\rmC}( x', x''' )G_2^{\rmC}( x'', x''' ) - G^\rmS_{2,1}(x,x';x''')\hat{G}_2^{\rmC}(x'', x''')
    \nonumber
    \\
    {}  - G^\rmS_{2,1}(x,x'';x''')\hat{G}_2^{\rmC}(x', x''') - G^\rmS_{2,1}(x',x''; x''')\hat{G}_2^{\rmC}(x, x''') \Big] - G^\rmS_3(x,x',x'').
    \label{eq:Cmexp}
\end{align}
\end{widetext}
It follows that 
\begin{equation}\label{eq:sourceCieven}
    \mathbb{D}_{\bar{\phi}} \hat{\phi}^{\Citer}_{(2)} = \frac{1}{2} \dddot{V}(\bar{\phi}) \big( \hat{\phi}_{(1)}^\rmC \big)^2,
\end{equation}
so $\hat{\phi}^{\Citer}$ is formally a solution to the homogeneous field equation \eqref{vacPhi}. These results are described as formal because the iterated time-even prescription is not expected to be well-defined when the retarded or the advanced Green's functions are nonzero inside their lightcones. Such (tail) effects can be interpreted as failures of Huygens' principle. They occur generically in curved spacetimes \cite{Guenther, introEMRI}, where fields scatter off the spacetime curvature and propagate into the light cone. They occur even for linear fields on flat spacetime when those fields are massive.

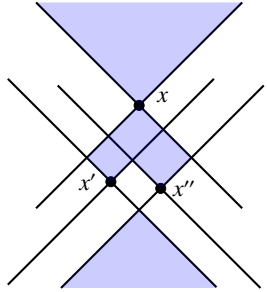
\begin{figure}[h!]
    \centering
    \begin{tikzpicture}[scale=2.2]

\def\b{0.9}

  \useasboundingbox (-\b,-\b) rectangle (\b,\b);

  \def\L{0.62}

  \def\xzero{0.2}
  \def\tzero{-0.2}
  \coordinate (x) at (\xzero,\tzero);
  \filldraw[black] (x) circle (0.03);
  \node at (\xzero+0.14,\tzero) {$x''$};

  \draw[thick] (x) -- ++(\L,\L);
  \draw[thick] (x) -- ++(-\L,\L);
  \draw[thick] (x) -- ++(0.6,-0.6);
  \draw[thick] (x) -- ++(-0.6,-0.6);

  \def\yzero{-0.1}
  \def\ttwo{-0.16}
  \coordinate (y) at (\yzero,\ttwo);
  \filldraw[black] (y) circle (0.03);
  \node[black] at (\yzero-0.14,\ttwo) {$x'$};

  \draw[thick] (y) -- ++(\L,\L);
  \draw[thick] (y) -- ++(-\L,\L);
  \draw[thick] (y) -- ++(0.635,-0.635);
  \draw[thick] (y) -- ++(-\L,-\L);

  \def\zzero{0.07}
  \def\tthree{0.3}
  \coordinate (z) at (\zzero,\tthree);
  \filldraw[black] (z) circle (0.03);
  \node[black] at (\zzero+0.14,\tthree+0.05) {$x$};

  \draw[thick] (z) -- ++(\L,\L);
  \draw[thick] (z) -- ++(-\L,\L);
  \draw[thick] (z) -- ++(\L,-\L);
  \draw[thick] (z) -- ++(-\L,-\L);


  \pgfmathsetmacro{\tBot}{0.5*(\xzero + \tzero - \yzero + \ttwo)}
  \pgfmathsetmacro{\xBot}{\xzero - (\tBot - \tzero)}

  \pgfmathsetmacro{\tRight}{0.5*(\zzero + \tthree - \yzero + \ttwo)}
  \pgfmathsetmacro{\xRight}{\yzero + (\tRight - \ttwo)}

  \pgfmathsetmacro{\tLeft}{0.5*(\xzero + \tzero - \zzero + \tthree)}
  \pgfmathsetmacro{\xLeft}{\xzero - (\tLeft - \tzero)}

  \pgfmathsetmacro{\tLL}{0.5*(\yzero + \ttwo - \zzero + \tthree)}
  \pgfmathsetmacro{\xLL}{\yzero - (\tLL - \ttwo)}

  \pgfmathsetmacro{\tRR}{0.5*(\zzero + \tthree - \xzero + \tzero)}
  \pgfmathsetmacro{\xRR}{\xzero + (\tRR - \tzero)}

  \fill[blue,opacity=0.2]
    (y) --
    (\xBot,\tBot) --
    (\xLeft,\tLeft) --
    (\xLL,\tLL) -- cycle;

  \fill[blue,opacity=0.2]
    (x) --
    (\xRR,\tRR) --
    (\xRight,\tRight) --
    (\xBot,\tBot) -- cycle;

  \fill[blue,opacity=0.2]
    (\xBot,\tBot) --
    (\xRight,\tRight) --
    (z) --
    (\xLeft,\tLeft) -- cycle;


  \def\tTop{0.92}

  \pgfmathsetmacro{\xTopLeft}{\zzero - (\tTop - \tthree)}
  \pgfmathsetmacro{\xTopRight}{\zzero + (\tTop - \tthree)}

  \fill[blue,opacity=0.2]
    (z) --
    (\xTopRight,\tTop) --
    (\xTopLeft,\tTop) -- cycle;


  \def\tBottom{-0.8}

  \pgfmathsetmacro{\tPastTop}{0.5*(-\xzero + \tzero + \yzero + \ttwo)}
  \pgfmathsetmacro{\xPastTop}{\yzero + (\ttwo - \tPastTop)}

  \pgfmathsetmacro{\xPastLeft}{\xzero - (\tzero - \tBottom)}
  \pgfmathsetmacro{\xPastRight}{\yzero + (\ttwo - \tBottom)}

  \fill[blue,opacity=0.2]
    (\xPastTop,\tPastTop) --
    (\xPastRight,\tBottom) --
    (\xPastLeft,\tBottom) -- cycle;

\end{tikzpicture}
    \caption{The conservative 3-point function $\hat{G}^{\Citer}_3(x,x',x'')$ depends on an integral of $G^\rmC_2(x,x''')G^\rmC_2(x''',x')G^\rmC_2(x''',x'')$ over all spacetime points $x'''$. The points where this integrand might be nonzero are illustrated here in blue. The region is unbounded in both the past and the future, which is due to the conservative 2-point function generically having support for all timelike-separated points.}
    \label{fig:domainforcons3point}
\end{figure}

When tails exist, the term in Eq.\ \eqref{eq:Cmexp} that involves three factors of $G_2^\rmC$ is an integral over all points that are causally connected to $x$, $x'$ and $x''$, as shown in Fig. \ref{fig:domainforcons3point}.
This region is unbounded, so computing $\hat{G}_3^{\Citer}$ generically requires integrating over an infinite spacetime volume. If the tails do not decay sufficiently rapidly, $\hat{G}_3^{\Citer}$ therefore fails to exist. We show in Appendix \ref{sec:appendixA} that the tails in a simple analytic example do not, in fact, decay fast enough to render the integral finite. That construction is not intended as a general proof that iterated time-even 3-point functions are divergent in arbitrary theories. Rather, it serves as a concrete illustration that such divergences can arise, and suggests that they may persist more broadly. We conclude that the iterated time-even sector cannot necessarily be expected to extend from first to second order. 

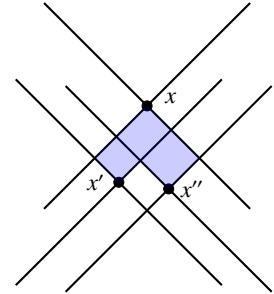
\begin{figure}[h!]
    \centering
    \begin{tikzpicture}[scale=2.2]

  \useasboundingbox (-1,-1) rectangle (1,1);

  \def\L{0.62}

  \def\xzero{0.2}
  \def\tzero{-0.2}
  \coordinate (x) at (\xzero,\tzero);
  \filldraw[black] (x) circle (0.03);
  \node at (\xzero+0.14,\tzero) {$x''$};

  \draw[thick] (x) -- ++(\L,\L);
  \draw[thick] (x) -- ++(-\L,\L);
  \draw[thick] (x) -- ++(\L,-\L);
  \draw[thick] (x) -- ++(-\L,-\L);

  \def\yzero{-0.1}
  \def\ttwo{-0.16}
  \coordinate (y) at (\yzero,\ttwo);
  \filldraw[black] (y) circle (0.03);
  \node[black] at (\yzero-0.14,\ttwo) {$x'$};

  \draw[thick] (y) -- ++(\L,\L);
  \draw[thick] (y) -- ++(-\L,\L);
  \draw[thick] (y) -- ++(\L,-\L);
  \draw[thick] (y) -- ++(-\L,-\L);

  \def\zzero{0.07}
  \def\tthree{0.3}
  \coordinate (z) at (\zzero,\tthree);
  \filldraw[black] (z) circle (0.03);
  \node[black] at (\zzero+0.14,\tthree+0.05) {$x$};

  \draw[thick] (z) -- ++(\L,\L);
  \draw[thick] (z) -- ++(-\L,\L);
  \draw[thick] (z) -- ++(\L,-\L);
  \draw[thick] (z) -- ++(-\L,-\L);


  \pgfmathsetmacro{\tBot}{0.5*(\xzero + \tzero - \yzero + \ttwo)}
  \pgfmathsetmacro{\xBot}{\xzero - (\tBot - \tzero)}

  \pgfmathsetmacro{\tRight}{0.5*(\zzero + \tthree - \yzero + \ttwo)}
  \pgfmathsetmacro{\xRight}{\yzero + (\tRight - \ttwo)}

  \pgfmathsetmacro{\tLeft}{0.5*(\xzero + \tzero - \zzero + \tthree)}
  \pgfmathsetmacro{\xLeft}{\xzero - (\tLeft - \tzero)}

  \pgfmathsetmacro{\tLL}{0.5*(\yzero + \ttwo - \zzero + \tthree)}
  \pgfmathsetmacro{\xLL}{\yzero - (\tLL - \ttwo)}

  \pgfmathsetmacro{\tRR}{0.5*(\zzero + \tthree - \xzero + \tzero)}
  \pgfmathsetmacro{\xRR}{\xzero + (\tRR - \tzero)}

  \fill[blue,opacity=0.2]
    (y) --
    (\xBot,\tBot) --
    (\xLeft,\tLeft) --
    (\xLL,\tLL) -- cycle;

  \fill[blue,opacity=0.2]
    (x) --
    (\xRR,\tRR) --
    (\xRight,\tRight) --
    (\xBot,\tBot) -- cycle;

  \fill[blue,opacity=0.2]
    (\xBot,\tBot) --
    (\xRight,\tRight) --
    (z) --
    (\xLeft,\tLeft) -- cycle;

\end{tikzpicture}
    \caption{The effective retarded 3-point function $\hat{G}^+_3(x,x',x'')$ is constructed, in part, by integrating $G^+_2(x,x''')G^+_2(x''',x')G^+_2(x''',x'')$ over all spacetime points $x'''$. The points where this integrand might be nonzero are illustrated here in blue. They form a bounded region that is a proper subset of the region illustrated in Fig. \ref{fig:domainforcons3point}.}
    \label{fig:domainforret3point}
\end{figure}

Identical integrals involving three factors of $G_2^\rmC$ also arise in Eqs.\ \eqref{eq:Ceexp} and \eqref{eq:Cpmexp}, which describe the 3-point functions associated with the (global) time-even and the symmetrized conservative sectors. Nevertheless, we do not expect similar infrared divergences to occur in those cases. The reason for this is that the term under discussion cannot be integrated on its own, but only in combination with other terms, and the full integral does not diverge. To see why this must be true, first note that $\hat{G}_3^{\rmC_\pm}$ is a symmetrization of $\hat{G}_3^\pm$. Recalling the definitions \eqref{G3plus} and \eqref{eq:retreg3point}, the only possibly-problematic parts of these latter 3-point functions would involve integrals of terms proportional to
\begin{equation}
    G_2^\pm(x,x''') G_2^\pm(x''',x') G_2^\pm(x''',x'').
\end{equation}
In the retarded case, the integration point $x'''$ must be simultaneously in the past of $x$ and in the future of both $x'$ and $x''$, as shown in Fig. \ref{fig:domainforret3point}.
This causal restriction always produces a finite domain of integration, suggesting that at least $\hat{G}_3^+$ must be infrared-finite. An almost identical consideration also applies to $\hat{G}_3^-$. Any symmetrization of a finite $n$-point function is also finite, so $\hat{G}_3^{\rmC_\pm}$ must therefore be finite.

The finiteness of the time-even 3-point function can instead be argued by noting that it may be written as
\begin{equation}
    \hat{G}_3^{\Ceven} = \frac{1}{2} \psym \big( \hat{G}^+_3 + \hat{G}^-_3 \big) .
\end{equation}
This is a symmetrization of two finite 3-point functions, so it too is  finite.

\subsection{Applicability to bound orbits}

We have just described an infrared divergence that is expected to arise in the 3-point function $\hat{G}^{\Citer}_3$. However, even if we use a different 3-point function that is well-defined, another type of divergence can arise when integrating that 3-point function to compute a field. Infrared divergences of this type are expected to arise for \textit{all} conservative sectors we consider, at least for bound orbits around a central charge at second order. These divergences require us to restrict attention to unbound (scattering) orbits in order to define finite conservative and dissipative
sectors\footnote{We thank Adam Pound for bringing this issue to 
our attention.}. 

In conservative dynamics, the absence of dissipation implies that for bound orbits,  worldlines are eternally bound. As a result, a binary remains strongly interacting at all times and continuously emits radiation. At first order, this poses no fundamental difficulty. Although the total integrated energy in the first-order field is infinite, the field itself remains finite in the neighborhood of the secondary. Consequently, the associated first-order self-force
is well defined. At second order, however, the situation is qualitatively different. The first-order field now acts as a source for the second-order field, and the second-order solution at a given spacetime point is obtained by integrating products of first-order fields over all spacetime. Due to the slow falloff of the first-order fields, these integrals are generically expected to diverge even when the second-order field is evaluated at a finite distance from the primary. Whether or not these divergences can be resolved in a useful way remains an open question.


\section{\label{sec:Hamiltonian} Hamiltonian formulations of
    conservative sectors}

We now construct Hamiltonians for the time-even and the symmetric
conservative dynamical systems defined in Sec.\ \ref{sec:cons2nd}
above, which is accomplished using the general result described in
Appendix \ref{sec:Hamproof}. That result is derived perturbatively in
a small parameter $\epsilon$ that is used to track the order of
nonlocal terms. However, the physical systems studied in this paper
have thus far been studied in a perturbative expansion in powers of
the scalar charge $\hat{q}$. As can be seen in
Eqs.\ (\ref{eq:hatphiexpr}) and (\ref{eq:scalareomeff}), the force has
a local term proportional to $\hat{q}$, a leading
potentially-nonlocal\footnote{We refer to this term as
potentially-nonlocal since, even though it takes the form of a
nonlocal integral, under certain circumstances this integral can
reduce to a local term.} term proportional to $\hat{q}^2$ and a
subleading potentially-nonlocal term proportional to $\hat{q}^3$.  In
Appendix \ref{sec:Hamproof}, both the leading and the subleading
potentially-nonlocal terms in the force are $O(\epsilon)$, and the
result is derived accurate to $O(\epsilon)$. This accuracy is sufficient for
our context here, since
the errors are quadratic $\epsilon$ and so are of order
$O(\hat{q}^4)$, which is one order smaller than the subleading
nonlocal term of order $O(\hat{q}^3)$ which we retain. We distinguish between these
two small parameters by using parentheses, such as $(n)$,  to denote
the order in $\hat{q}$, while using square brackets, such as $[n]$, to
denote the order in $\epsilon$. Furthermore, we use $(\le n)$ or $[\le
  n]$ to denote a quantity that includes all orders up to and including
$n$.

The derivation in Appendix \ref{sec:Hamproof} begins with a nonlocal action principle of the form (\ref{eq:action}). The symmetry of the $n$-point functions constructed in the previous subsections implies that the dynamics can be expressed in terms of such an action, where the zeroth-order Hamiltonian appearing in that action is given by
\be
\label{h0def}
H^{[0]}=-\sqrt{-g^{\mu\nu}p_\mu p_\nu}-\hat{q}\bar{\phi}.
\ee
Note that this Hamiltonian is denoted by a superscript $[0]$ since it
is zeroth order in any potentially-nonlocal terms, that is, zeroth
order in $\epsilon$, see Eq.\ (\ref{eq:hexpand}) below.
We denote by $X^{[0]}_{\tau}(Q)$ the corresponding flow, where $\tau$
is proper time. We can also expand this flow in powers of ${\hat q}$.
We denote by $X^{(0)}_{\tau}(Q)$ the flow at zeroth order in
$\hat{q}$, and by $X^{(\leq 1)}_\tau(Q)$ the flow that takes
into account both the zeroth and the first-order contributions from
$\hat{q}$.

Besides the purely local part of the action that involves $H^{[0]}$, there is also a potentially-nonlocal part with the form\footnote{We assume that despite the singular nature of $\hat{G}_2^{\rmC}$ and $\hat{G}_3^{\rmC}$, this action can be at least formally manipulated. This is similar to our earlier assumption that Eq.\ (\ref{eq:hatphiexpr}) is well-defined.} 
\begin{align}\label{eq:nonlocalactionscalar}
S_{\text{nl}}[z,p]&=-\frac{\hat{q}^2}{2}\int \! \rmd\tau \rmd\tau' \hat{G}_2^{\rmC} (z_\tau,z_{\tau'})
\nonumber
\\
&-\frac{\hat{q}^3}{3} \int \! \rmd\tau \rmd\tau' \rmd\tau''
\hat{G}_3^{\rmC}(z_\tau,z_{\tau'},z_{\tau''}) + O({\hat q}^4).
\end{align}
Here, $\hat{G}^{\rmC}_2$ is the conservative 2-point function (\ref{eq:CD2point}) that is associated with the first-order effective field. The 3-point function 
$\hat{G}^{\rmC}_3$ can be either of the expressions 
(\ref{eq:Ceexp}) or (\ref{eq:Cpmexp}), depending on whether we adopt
the time-even or the symmetrized conservative sectors.
One can check that the equation of motion obtained from Eqs.\ \eqref{h0def} and \eqref{eq:nonlocalactionscalar} coincides with that derived in Sec. \ref{sec:cons2nd} given by Eqs.\ \eqref{eq:scalareomeff} and \eqref{eq:hatphi2C}.
The potentially-nonlocal action \eqref{eq:nonlocalactionscalar} has the same form as the nonlocal
action \eqref{eq:nonlocalaction}, where the $n$-point functions $\epsilon \mathcal{G}_2$ and $\epsilon \mathcal{G}_3$ in that latter expression are replaced by $\hat{q}^2 \hat{G}^\rmC_2$ and $\hat{q}^3 \hat{G}^\rmC_3$, respectively. The time parameter $s$ used in the Appendix has also been replaced here with the proper time $\tau$.

With these replacements, the Hamiltonian is given by
Eq.\ (\ref{eq:hamsecondorder}), up to corrections which are quadratic
in the potentially-nonlocal terms (and therefore of order
$\hat{q}^4$).
From Eqs.\ \eqref{h0def}, \eqref{eq:nonlocalactionscalar},
\eqref{eq:nonlocalaction} and \eqref{eq:hamsecondorder} we find
\be\label{eq:hamsecondorderscalar}
\begin{aligned}
{\check H}^{[\le1]}(Q)&=-\sqrt{-g^{\mu\nu} p_\mu p_\nu}-\hat{q}
\bar{\phi}(Q)+\frac{\hat{q}^2}{2}\int \! \rmd\tau
\, \hat{G}^{\rmC}_2 \big[ Q,X^{[0]}_\tau(Q)\big] 
\\
& {} +\frac{\hat{q}^3}{2}\int
\! \rmd\tau \rmd\tau'\,
\hat{G}_3^{\rmC} \big[ Q,X^{[0]}_\tau(
    Q),X^{[0]}_{\tau'}(Q)\big] + O({\hat q}^4).
\end{aligned}
\ee
Since we are expanding in powers of ${\hat q}$, we can replace
the flow $X^{[0]}_\tau$ in the last term with $X^{(0)}_\tau$, the flow to
zeroth order in ${\hat q}$, and similarly replace
$X^{[0]}_\tau$ in the second last last term with $X^{(\le1)}_\tau$. In
both cases the replacement incurs an error of order $O({\hat q}^4)$.
This yields
\be\label{eq:hamsecondorderscalar1}
\begin{aligned}
{\check H}^{(\le3)}(Q)&=-\sqrt{-g^{\mu\nu} p_\mu p_\nu}-\hat{q}
\bar{\phi}(Q)+\frac{\hat{q}^2}{2}\int \! \rmd\tau
\, \hat{G}^{\rmC}_2 \big[ Q,X^{(\le1)}_\tau(Q)\big] 
\\
& {} +\frac{\hat{q}^3}{2}\int
\! \rmd\tau \rmd\tau'\,
\hat{G}_3^{\rmC} \big[ Q,X^{(0)}_\tau(
    Q),X^{(0)}_{\tau'}(Q)\big].
\end{aligned}
\ee
Here the 3-point function $\hat{G}_3^{\rmC}$ can again be either of
the expressions (\ref{eq:Ceexp}) or (\ref{eq:Cpmexp}). The
corresponding symplectic
form ${\check \Omega}^{(\le3)}$ is given by Eqs.\ \eqref{Omegacan} and
\eqref{eq:Omegasimp}, and therefore
takes the standard form in the coordinates $Q^A = (p_\mu, x^\mu)$. 
Finally we note that the flow ${\check X}_\tau$ associated with this
Hamiltonian
is related to the physical flow $X_\tau$ by the pullback \eqref{eq:pullback1}
generated by the
linearized phase space diffeomorphism \eqref{zetadef}, which by similar arguments reduces here to
\begin{widetext}
\begin{align}
  \label{zetadef1}
  \zeta^A(Q)  = - \frac{1}{2} \Omega^{[0]\,AB} & \frac{\partial}{\partial
    Q^B} \left\{ {\hat q}^2 \int \! \rmd \tau \rmd \tau' \, \chi_2(\tau,\tau')
\, \hat{G}^{\rmC}_2 \big[ X^{(\le1)}_\tau(Q),
    X^{(\le1)}_{\tau'}(Q') \big] 
    \right. \nonumber \\  & {} 
  + {\hat q}^3 \int \! \rmd \tau \rmd \tau' \rmd \tau'' \chi_3(\tau,\tau',\tau'') 
  \hat{G}^{\rmC}_3 \big[ X^{(0)}_\tau(Q),
    X^{(0)}_{\tau'}(Q'), X^{(0)}_{\tau''}(Q'') \big]  \bigg\}
  \bigg|_{Q'' = Q' = Q} + O({\hat q}^4).
  \end{align}
\end{widetext}

\section{\label{sec:conclusions} Discussion}

This paper has investigated different definitions for conservative sectors of the second-order scalar self-force. We have focused on definitions that admit Hamiltonian descriptions, applying a recently developed method for deriving local Hamiltonian descriptions of nonlocal-in-time dynamical systems to the case of the second-order scalar self-force. The starting point of this method is a nonlocal action principle, expressed in terms of integrals involving fully-symmetric $n$-point functions. To cast the self-force problem into this form, we applied a technique based on field transformations that maps the physical field to a source-free effective field. These transformations effectively absorb the contribution of the singular self-field into a redefinition -- or ``dressing'' -- of an object's charge density and stress-energy tensor. 

Although all natural definitions for the first-order conservative self-force agree with one another, definitions begin to differ at second order. We have described three types of conservative sector for the second-order self-force, each of which admits a Hamiltonian description: the iterated time-even, the time-even, and the symmetric sectors. While the first of these sectors is not expected to be viable, all three definitions are summarized in Table \ref{tab:Prescriptions}. Each can be viewed as having been obtained from the full self-force by applying suitable combinations of the three projection operators defined in Sec. \ref{sec:projection}. One of these operators simply symmetrizes the $n$-point functions;  the remaining two provide different ways to extract time-even components. 

\renewcommand{\arraystretch}{1.2}%
\begin{table}
\setlength{\tabcolsep}{8pt}
  \begin{tabular}{l c  l  c}
   \hline\hline
   Prescription & Abbreviation & Definition &  $\hat{G}_3^{\cdots}$
   \\
    \hline
    Symmetric 	&	$\rmC_\pm$	&	$\psym \hat{\phi}^\pm$ 	& 	\eqref{eq:Cpmexp}	
    \\
    Time-even 	&	$\Ceven$	&  $\psym \pglob \hat{\phi}^\pm$	&  	\eqref{eq:Ceexp} 
    \\	
    Iterated time-even	&  $\Citer$	&   $\psym \piter \hat{\phi}^\pm$ 	&   \eqref{eq:Cmexp}\\
    \hline\hline
	\end{tabular}
	
    \caption{Summary of our conservative sectors. Definitions for the fields that are to be inserted into the force are written in terms of the projectors $\psym$, $\pglob$ and $\piter$ introduced in Sec. \ref{sec:projection}. These projectors act on the effective fields $\hat{\phi}^\pm$ that are associated with the retarded field $\phi^+$ or the advanced field $\phi^-$. In the symmetric case, the retarded and the advanced effective fields result in different projections. In the time-even and iterated time-even cases, these fields have the same projection. The final column here references the 3-point function that can be integrated to yield the second-order effective field for the given prescription. Although we do not expect the iterated time-even prescription to yield a finite 3-point function, it is included for completeness.} 
    
    \label{tab:Prescriptions}	
\end{table}

\subsection{Does a preferred prescription exist?}

It is natural to ask if there is any reason to prefer one conservative sector over another. Although we can discard the iterated time-even sector as being infrared-divergent, we see no compelling physical arguments for preferring the time-even prescription over the symmetric prescriptions. This lack of a unique prescription does not, however, have any physical consequences: only the total self-force is physical, and different prescriptions merely reshuffle terms between the conservative and the dissipative sectors of the splitting. The choice of prescription is therefore a pragmatic issue rather than an issue of principle. Its utility depends on the intended application.

One general comment that can nevertheless be made is that our conservative sectors can be distinguished by the sources that arise in the field equations satisfied by the various second-order fields, which are given by Eqs.\ (\ref{eq:sourceCeven}), (\ref{eq:sourceCpm}) and (\ref{eq:sourceCieven}). The iterative time-even prescription formally produces a second-order conservative field that is sourced by the product of two first-order conservative fields. The time-even prescription includes this source as well as the product of two first-order dissipative fields, which is also time-even. In addition to these terms, the source for the symmetric prescription also includes the (time-odd) product of the first-order conservative field and the first-order dissipative field. These definitions thus form a hierarchy of increasingly inclusive prescriptions, in the sense that each choice of conservative dynamics includes more effects than the one before. The symmetrized definition based on the retarded solution might thus be preferred on the basis that it includes as many of the physical effects as possible within the definition of the conservative sector. In particular, it includes conservative effects deriving from terms quadratic in the first-order dissipative field. These quadratic terms give rise to corrections to the conserved quantities of orbital motion.

This argument becomes clearer when computing gravitational waveforms for extreme mass-ratio inspirals at post-1-adiabatic order for LISA. One then needs only the phase-space average of the time derivatives of the energy, angular momentum, and Carter constant induced by the second-order self-force \cite{flanagan2}. It can be shown that the contribution from a fully symmetric 3-point function vanishes under this phase-space averaging, by properties of Hamiltonian dynamics\footnote{With the exception of resonant orbits.} \cite{Mino:2003yg,flanagan2}, so the phase-space averaged result depends only on the dissipative piece of the second-order self-force. In this context, the most efficient choice is therefore the symmetrized prescription, since it will yield the dissipative self-force with the fewest number of terms. Nevertheless, the time-even prescription could be used and would yield the same physical result, with some additional terms in the dissipative sector that vanish after averaging.

We thus suggest the symmetrized conservative sector as sometimes having a calculational advantage over the time-even sector, although the latter might be preferred in other contexts. Note that the symmetrized conservative sector is not time-even. Although this might be considered unusual for conservative systems, contributions to conservative dynamics that are not time-even can appear through, e.g., ``Schott terms;'' see Ref. \cite{Trestini:2025nzr} and references therein.

\subsection{Conservative and dissipative sectors in other approximations}

Conservative and dissipative sectors have been defined in various ways within other perturbative approaches to the two-body problem. We now discuss some of those approaches, focusing in particular on the post-Newtonian (PN) and the post-Minkowskian (PM) frameworks and their comparison with self-force (SF) expansions. Different physical effects arise at different orders in each of these expansions. For instance, effects of gravitational-wave emission on motion first appear at 2.5PN, 3PM and 1SF orders \cite{poissonwill,pound}. Similarly, the first occurrences of nonlocal interactions (commonly referred to as “tail” or “hereditary” effects) are typically said to appear at 4PN, 4PM, and 1SF orders \cite{Dlapa_2022,Damour_2014,introEMRI}.

As discussed in Appendix \ref{sec:Hamproof}, nonlocalities can obscure the identification of conservative and dissipative sectors of the dynamics, since the former are typically characterized by the existence of a local Hamiltonian on phase space. In other words, two Hamiltonian systems starting at the same point on phase space should follow the same trajectory. Nonlocal effects, however, make the system ``remember" its past trajectory, making them manifestly not Hamiltonian. A variety of techniques were used within each approximation scheme to ``localize" the dynamics, typically via order-reduction methods. Then, a unique conservative sector was obtained for the local dynamics at 4PN, 4PM and 1SF orders. At 4PN order, the conservative dynamics were first derived in \cite{Damour_2014,Damour_2015}. At 1SF order, they were obtained in \cite{tanaka, Blanco:2022mgd}. At 4PM order, the conservative sector was first obtained from scattering amplitudes in \cite{Bern_2022_PRL,Bern_2022_PoS}. All prescriptions for choosing the conservative sector agree at leading order in the nonlocalities. This is analogous to what happens for the linear self-force, as discussed in Section \ref{sec:cons1st}.

At the next order beyond the first appearance of nonlocalities, the dynamics exhibit a variety of more intricate nonlocal structures, including effects known as “failed tails,” ``radiation-reaction squared," “memory” and “tail-of-tail;” see Ref. \cite{porto2025nonlineargravitationalradiationreaction} and references therein for details about each effect. Since this terminology varies across different communities, we instead refer to these effects uniformly as \emph{next-to-leading nonlocalities}. The increased complexity at these orders has led to a wide range of methods \cite{Bini:2020nsb, Almeida_2021, Almeida_2023, porto2025nonlineargravitationalradiationreaction, Bern_2026_PRL, driesse2026conservativeblackholescattering}, including effective field theory and amplitude-based techniques. These approaches can, in principle, be cross-checked by performing double expansions (e.g., in both PN and PM schemes) and comparing gauge-invariant observables. Such comparisons have been successfully carried out at the level of leading order nonlocal effects\footnote{See, for example, Ref. \cite{porto2025nonlineargravitationalradiationreaction} for a crosscheck between 5PN and 4PM results. Note that a check at the level of next-to-leading order in nonlocalites would have to be simultaneously 5PN and 5PM. This is also true of work such as Ref. \cite{Bini_Damour_2025_tuttifrutti}, where a  crosscheck between 6PN and  5PM orders is done, but where the 5PM dynamics is only known up to 1SF order.}. However, at the time of writing, no analogous comparisons have confirmed agreement at next-to-leading order in the nonlocal sector. In fact, recent work in the PM framework has encountered divergent results and called for a need to clarify the definition of the conservative sector; see the conclusions in Ref. \cite{driesse2026conservativeblackholescattering}.

It lies beyond the scope of this paper to determine whether the various prescriptions adopted in the literature can be systematically recast in terms of the projectors introduced in Section \ref{sec:projection}. Nonetheless, we expect that the framework developed here may help clarify the origin of the ambiguities currently affecting this area of research.

\subsection{Future directions}

A key assumption underlying our formulation is the existence of a singular 3-point function $G^\rmS_3$ obeying certain properties outlined in Sec. \ref{sec:2ndOrder}. The existence of this 3-point function remains to be established, and is left as an issue for future investigation. Other directions for future research include addressing the infrared divergences associated with bound orbits, discussed in Sec.\ \ref{sec:cons2nd}, and generalizing the framework to the case of the gravitational self-force.

\vspace{0.1cm}
\noindent \textit{Acknowledgments:} This
research was supported in part by NSF grants PHY-2110463 and
NSF-2409350. We thank Adam Pound for helpful and detailed discussions on many aspects of this work, including infrared divergences and the second-order self-force.  We also thank Leor Barack for helpful discussions, and Raj Patil for clarifying the study of conservative dynamics in the PN and PM expansions.

\color{black}


\twocolumngrid
\bibliography{Ref.bib}

\appendix

\section{Notation}
\label{App:notation}

We consider a number of different scalar fields -- typically denoted by a variant of $\phi$ -- and also $n$-point functions, which are typically denoted by a variant of $G$. This appendix explains the various adornments we apply to those base symbols. First, superscripts applied to fields and $n$-point functions are summarized in Table \ref{Table:Notation}. Furthermore:
\begin{itemize}
    \item The subscript $n$ on $G_n$ denotes that this is a function that depends on $n$ points. The exception is $G_{2,1}^\rmS$, which is a 3-point function constructed by taking one functional derivative of the 2-point function $G_2^\rmS$; cf. Eq.\ \eqref{G21}.

    \item A field $\phi_{(n)}$ is the $n$th-order contribution to $\phi$, where orders are counted in powers of a body's net effective charge $\hat{q}$. 
    
    \item Hats can be applied to both fields and $n$-point functions. A field $\hat{\phi}$ is an effective field that appears in the equations of motion. It is obtained by appropriately transforming a corresponding physical field $\phi$. A hatted $n$-point function $\hat{G}_n$ is integrated against a source to construct $\hat{\phi}$.

    \item We construct perturbations on top of the background field $\bar{\phi}$.
\end{itemize}
For example, $\hat{\phi}_{(2)}^+$ denotes the second-order contribution to the effective field that is associated with a retarded physical field. Lastly, note that all $n$-point functions we consider can be functionals of a field. Except where indicated explicitly [as in, e.g., Eq.\ \eqref{Wexpand}], that field is always assumed to be the background $\bar{\phi}$.

\begin{table}[h]
  \centering
  \setlength{\tabcolsep}{9pt}

  \begin{tabular}{c l l}
    \hline\hline
    Superscript & Description & Equations \\[.0ex]
    \hline
    $+$, $-$	& Retarded, advanced 	& \eqref{eq:ret2pointfieldeq}, 
    \eqref{G3plus},
    \eqref{phiPlus}
    \\
    $\rmS$      & Singular  &  
    \eqref{Wexpand},
    \eqref{phiHatexpand},
    \eqref{eq:singularnpointeqs}
    \\
    $\rmC$      & Conservative & 
    \eqref{eq:CD2pointa},
    \eqref{eq:CD2point},
    \eqref{phiCHat1}
    \\
    $\rmD$      & Dissipative &
    \eqref{eq:CD2pointa},
    \eqref{phiD}
    \\
    $\Ceven$    & Time-even cons. & 
    \eqref{phiEven},
    \eqref{eq:Ce},
    \eqref{eq:Ceexp}
    \\
    $\Citer$    & Iterated time-even cons. & 
    \eqref{phiIeven},
    \eqref{eq:Cm},
    \eqref{eq:Cmexp}
    \\
    $\rmC_\pm$  & Symmetrized cons. & 
    \eqref{phiSym},
    \eqref{eq:Cpm},
    \eqref{eq:Cpmexp}
    \\[.2ex]
    \hline\hline
  \end{tabular}
  
  \caption{Table of superscripts applied to fields and $n$-point functions. The final three rows, which refer to definitions for conservative sectors that differ at second order, are summarized differently in Table \ref{tab:Prescriptions} above.}
  \label{Table:Notation}
\end{table}

\section{Local Hamiltonian dynamics from nonlocal-in-time action principles}

\label{sec:Hamproof}

This Appendix reviews the results of Ref.\ \cite{blanco2024localhamiltoniandynamicsnonlocal}, where one of
the authors showed that a certain class of dynamical systems with perturbative, nonlocal-in-time interactions are Hamiltonian order by order in perturbation theory.  We first define the relevant class of systems, then describe the local-in-time systems obtained by treating the nonlocalities perturbatively and finally give local Hamiltonian descriptions of these systems. This formalism is applied in Sec. \ref{sec:Hamiltonian} above to find Hamiltonians for conservative sectors of the second-order scalar self-force problem.

\subsection{General class of nonlocal dynamical systems}
\label{sec:nonlocalactions} 

The equations of motion we consider are
integro-differential equations, as opposed to the ordinary
differential equations characteristic of simpler local-in-time
systems. An example of such an integro-differential equation is
\be\label{eq:nonlocalexample}
\ddot{x}(t)=f(x(t))+\int_{-\infty}^{\infty} \! \rmd t' K(t,t')x(t').
\ee
Here, $f(x)$ is proportional to the local piece of the force and the integral is a nonlocal-in-time force that is a functional of the path $x(t')$. The
2-point function $K(t,t')$ captures the nonlocal interaction.
In cases where this 2-point function is symmetric, the nonlocal dynamics of Eq.\ (\ref{eq:nonlocalexample}) can be obtained from a term of the form
\be\label{eq:nonlocalactionexample}
S_{\rm nl}[x]=\frac{1}{2}\int \! \rmd t \rmd t' K(t,t')x(t)x(t')
\ee
in the action functional.  Note that this term depends only on the
symmetric piece of $K(t,t')$, because of the double integral over
time.  If $K(t,t')$ is not
symmetric, there is no action principle formulation\footnote{Unless
  one doubles the number variables, as in the Schwinger-Keldysh
  formalism \cite{1961JMP.....2..407S,1360865820894123264, Galley:2012hx}.} and the dynamics are not Hamiltonian. We are interested in systems that do admit Hamiltonian formulations, and we shall see that such formulations exist whenever the dynamics can be written in terms of suitably-symmetric $n$-point functions.

In this appendix, we consider a phase space $\Gamma$ parametrized by a coordinate system
$Q^A=(x^\mu,p_\mu)$ and action functionals of paths $X_s=\big(x^\mu(s),p_\mu(s)\big)$ of the form \be\label{eq:action}
S[X]=\int \! \rmd s  \dot{x}^\mu p_\mu-\int \! \rmd s H^{[0]}(X_s) + S_{\rm nl}[X],
\ee
where $H^{[0]} (Q)$ is a local Hamiltonian function on phase space. Generalizing Eq.\ \eqref{eq:nonlocalactionexample}, the nonlocal piece of the action appearing here is assumed to have the form 
\begin{eqnarray}
\label{eq:nonlocalaction}
  S_{\rm nl}[X]&=&
  - \frac{\epsilon}{2}\int \! \rmd s_1  \rmd s_2 \mathcal{G}_2(X_{s_1},X_{s_2}) \nonumber \\
  &&  {} - \frac{\epsilon}{3}\int \! \rmd s_1 \rmd s_2 \rmd s_3 \mathcal{G}_3(X_{s_1},X_{s_2},X_{s_3}),
\end{eqnarray}
where $\mathcal{G}_2$ and $\mathcal{G}_3$ are some 2-point and 3-point
functions and $\epsilon$ is a formal expansion parameter. This expansion parameter is used for order counting in the computation of the Hamiltonian and the symplectic form. It differs from the order counting in powers of the scalar charge used in the majority of this paper, where $\mathcal{G}_2$ and $\mathcal{G}_3$ are associated with first- and second-order effects, respectively. Regardless, the nonlocal action here depends only on the fully-symmetric pieces of $\mathcal{G}_2$ and $\mathcal{G}_3$. Without loss of generality, we therefore assume that they are fully symmetric. We also assume that these functions satisfy appropriate asymptotic fall-off conditions, detailed in
Ref.\ \cite{blanco2024localhamiltoniandynamicsnonlocal}.

Dynamical systems described by the actions (\ref{eq:action}) are a specialization of those analyzed in Ref.\ \cite{blanco2024localhamiltoniandynamicsnonlocal}, in three ways: the $n$-point functions here do not have any explicit dependence on time, we retain only two $n$-point functions, ${\mathcal G}_2$ and ${\mathcal G}_3$, and we  work only to linear order in $\epsilon$ rather than to all orders. However, this special case is sufficient for the application of this paper\footnote{Applying this formalism for the second-order gravitational (rather than the scalar) self-force requires use of the results of
Ref.\ \cite{blanco2024localhamiltoniandynamicsnonlocal} up to $O(\epsilon^2)$, rather than up to $O(\epsilon)$.}.
The conservative sectors we define in Sec.\ \ref{sec:Hamiltonian} have actions of the form \eqref{eq:nonlocalaction}, and therefore can be written in Hamiltonian form. Essentially, we only need to guarantee that the dynamics are entirely determined by appropriately symmetric $n$-point functions.

The equations of motion for the dynamical systems \eqref{eq:action} can be
written as follows.
We first define the symplectic form
\be
\label{Omegacan}
\Omega^{[0]} \equiv \rmd p_\mu \wedge \rmd x^\mu.
\ee
Also define a flow on phase space to be a mapping 
\be
X:\mathbb{R}\times\Gamma\rightarrow \Gamma : (s, Q) \to X_s(Q),
\ee
which takes any point $Q\in\Gamma$ into $X_s(Q)\in \Gamma$. The flow is required to be the identity map at $s=0$, meaning that
\be
X_0(Q)=Q,
\ee
and also to satisfy the composition rule
\be\label{eq:propcomp}
X_s(X_{s'}(Q))=X_{s+s'}(Q)
\ee
for all $s,s'\in \mathbb{R}$.

It is now convenient to write the equations of motion in terms of a quantity $\Phi(Q,Q';X]$, which is a function of two points in phase space, $Q$ and $Q'$, and a functional of an arbitrary flow $X$ on $\Gamma$. It is given by
\begin{align}
  \label{eq:phi}
  \Phi(Q,Q'; X] &\equiv
  \epsilon \int \! \rmd s_1 \
  \mathcal{G}_2 \big( Q,X_{s_1}(Q') \big)
  \nonumber
  \\
 &{} +  \epsilon \int \! \rmd s_1 \rmd s_2 \ \mathcal{G}_3 \big( Q,X_{s_1}(Q'),X_{s_2}(Q') \big).
\end{align}
The equations of motion obtained by varying the action functional
(\ref{eq:action}) with respect to the trajectory $X_s$ are 
\be\label{eq:hameqs}
\Omega^{[0]}_{AB}\left.\frac{\rmd X^B_s}{\rmd s}\right.=\left[\frac{\partial}{\partial Q^A}H^{[0]}(Q)+\frac{\partial}{\partial Q^A}\Phi(Q,Q';X]\right]_{Q'=Q=X_s}.
\ee
The subscript $Q'=Q = X_s$ here means that first two arguments of $\Phi$ are evaluated at coincidence after differentiating $\Phi$ with respect to its first entry, and that subsequently the whole right hand side is evaluated at $X_s$. We also note that since $X_s$ is a flow, it will be a function of some initial condition $\bar{Q}$ on both sides of the equation.

\subsection{Local dynamical systems obtained by treating nonlocalities perturbatively}
\label{sec:localbyperturb}

Equation (\ref{eq:hameqs}) is an integro-differential system of
equations for the trajectories $X_s$ on phase space. Because of this, solutions are generally not parametrized by initial data consisting of a single phase space point $Q$. In fact, the space of initial data required to determine solutions of integro-differential systems of equations can be infinite dimensional and require derivatives of $x^\mu$ and $p_\mu$ with respect to time of all orders \cite{llosa1994hamiltonian}.

However, if we take the nonlocal contribution to the action to be small, the problem can be treated perturbatively in $\epsilon$ using a procedure called reduction of order. We first define the zeroth-order flow $X^{[0]}_s(Q)$ to be that generated by the Hamiltonian $H^{[0]}$, dropping all terms in Eq.\ (\ref{eq:hameqs}) due to $S_{\rm nl}$.
The first order flow can be written as 
\be
X^{[\le1]}_s(Q) = X^{[0]}_s(Q) + \epsilon X^{[1]}_s(Q).
\ee
The functional $\Phi$ in equation (\ref{eq:hameqs}) can then be evaluated
on the zeroth-order flow, defining the first order flow
$X^{[\le 1]}_s$ to satisfy\footnotemark
\begin{equation}\label{eq:firstorderflow}
\Omega^{[0]}_{AB}\frac{\mathrm{d} X_s^{[\le 1]B}(Q)}{\mathrm{d}s}
=
v_A \big( X^{[\le 1]}_s(Q) \big),
\end{equation}
\footnotetext{As described in Sec.\ \ref{sec:Hamiltonian}, a superscript $[n]$ denotes a term of order $O(\epsilon^n)$,
whereas a superscript $[\le n]$ denotes a term that includes all orders up to
$O(\epsilon^n)$. Note that this is different from the notation used in
Ref.~\cite{blanco2024localhamiltoniandynamicsnonlocal}, where square brackets
have a different meaning. We chose this notation to faciliate the application of the results in Sec.\ \ref{sec:Hamiltonian}.}
where 
\begin{align}
\label{eq:firstorderflow1}
v_A(Q)  
\equiv \frac{\partial}{\partial Q^A}H^{[0]} (Q) +\left[\frac{\partial}{\partial Q^A}\Phi(Q,Q';X^{[0]}]\right]_{Q'=Q}.
\end{align}
This forms a system of ordinary differential equations, local in time, which determine the first-order flow $X^{[\le 1]}$ once the zeroth-order flow $X^{[0]}$ is known.

\subsection{Local Hamiltonian description}

Although the perturbative expansion of subsection \ref{sec:localbyperturb}
alleviates the problem of nonlocality, the equations of motion 
(\ref{eq:firstorderflow}) are not manifestly Hamiltonian.  In the second term on the right hand side of Eq.\ (\ref{eq:firstorderflow1}), the derivative with respect to the first argument evaluated at coincidence is not necessarily of the form of a total derivative, so there is no obvious candidate for a Hamiltonian.

Nevertheless, dynamical systems described by action principles of
the form (\ref{eq:nonlocalaction}) do admit local Hamiltonian
descriptions to any finite order in $\epsilon$ \cite{blanco2024localhamiltoniandynamicsnonlocal}.
At first order, this means that there exists a Hamiltonian function
\be
\label{eq:hexpand}
   H^{[\le1]} = H^{[0]} + \epsilon H^{[1]}
   \ee
   and a symplectic form
\be
   \Omega_{AB}^{[\le1]} = \Omega_{AB}^{[0]} + \epsilon \Omega_{AB}^{[1]}
   \ee
   for which the equation of motion \eqref{eq:firstorderflow1} coincides with the Hamiltonian equation of motion to linear order
   \begin{equation}
     \label{eq:ham11}
   \Omega^{[\le 1]}_{AB} \frac{ \rmd{X}_s^{[\le1]B} }{\rmd s} =
   \frac{\partial H^{[\le1]}}{\partial Q^A} + O(\epsilon^2).
\end{equation}

The results for the Hamiltonian and symplectic form can be simplifed by using a diffeomorphism on phase space.
If we define a linearized diffeomorphism
\be
\label{deformed}
\psi : \Gamma \to \Gamma: Q^A \to Q^A + \epsilon \zeta^A(Q) + O(\epsilon^2),
\ee
for some vector field $\zeta^A$, we can define a new Hamiltonian and a
new symplectic form by taking
pullbacks, ${\check H} = \psi_* H$ and ${\check \Omega}_{AB} = \psi_* \Omega_{AB}$.  This yields
\begin{subequations}
  \label{eq:hamnew}
  \begin{eqnarray}
    {\check H}^{[\le1]} &=& H^{[0]} + \epsilon {\check H}^{[1]}, \\
    {\check \Omega}^{[\le1]}_{AB} &=& \Omega^{[0]}_{AB} + \epsilon {\check
      \Omega}_{AB}^{[1]}, 
  \end{eqnarray}
  \end{subequations}
where
\begin{subequations}
  \begin{eqnarray}
    {\check H}^{[1]} &=& H^{[1]} + \lie_\zeta H^{[0]}, \\
    {\check \Omega}^{[1]}_{AB} &=& \Omega^{[1]}_{AB} + \lie_\zeta \Omega_{AB}^{[0]}.
  \end{eqnarray}
\end{subequations}
We choose the vector field to make ${\check \Omega}_{AB}^{[1]} =0$, so
that
\be
\label{eq:Omegasimp}
   {\check \Omega}^{[\le1]}_{AB} = \Omega^{[0]}_{AB}.
   \ee
The integral curves
\be
   {\check X}^{[\le1]\,A}_s = {\check X}^{[0]\,A}_s + \epsilon {\check X}^{[1]\,A}_s
\ee
   of the new Hamiltonian system \eqref{eq:hamnew} are related to the solutions of $X^A_s$
the original equation of motion \eqref{eq:firstorderflow1} by the
pullback:
\be
\label{eq:pullback1}
   {\check X}_s^{[0]} = X_s^{[0]}, \ \ \ \
   {\check X}_s^{[1]} = X_s^{[1]} - \zeta^A(X^{[0]}_s).
\ee
These integral curves are given by a checked version of
Eq.\ \eqref{eq:ham11} with the simplification \eqref{eq:Omegasimp} to the symplectic form:   
\begin{equation}
   \Omega^{[0]}_{AB} \frac{ \rmd{{\check X}}_s^{[\le1]B} }{\rmd s} =
   \frac{\partial {\check H}^{[\le1]}}{\partial {Q}^A} + O(\epsilon^2).
\end{equation}
Note that these equations have the standard canonical form in the coordinates $Q^A =
(p_\mu,x^\mu)$, by Eq.\ \eqref{Omegacan}.

The explicit results for the diffeomorphism and Hamiltonian are given in
Eqs.\ (20) and (47) of
Ref.\ \cite{blanco2024localhamiltoniandynamicsnonlocal} and are as
follows.  The vector field $\zeta^A$ on $\Gamma$ is given by
\begin{widetext}
\begin{align}
  \label{zetadef}
  \zeta^A(Q)  = - \frac{1}{2} \Omega^{[0]\,AB} & \frac{\partial}{\partial
    Q^B} \left\{ \int \! \rmd s \rmd s' \chi_2(s,s') \mathcal{G}_2 \big[ X^{[0]}_s(Q),
    X^{[0]}_{s'}(Q') \big] 
    \right. \nonumber \\  & {} 
  + \int \! \rmd s \rmd s' \rmd s'' \chi_3(s,s',s'') 
  \mathcal{G}_3 \big[ X^{[0]}_s(Q),
    X^{[0]}_{s'}(Q'), X^{[0]}_{s''}(Q'') \big]  \bigg\}
  \bigg|_{Q'' = Q' = Q}.
  \end{align}
Here the derivative is taken before the evaluation at coincidence,
$X^{[0]}_s$ is the flow associated with the zeroth-order 
Hamiltonian $H^{[0]}$, and we have defined the functions
\begin{subequations}
\begin{eqnarray}  
  \chi_2(s,s') &\equiv& \frac{1}{2}[ \rm{sgn}(s) - \rm{sgn}(s')], \\
  \chi_3(s,s',s'') &\equiv& \frac{1}{2}[ \rm{sgn}(s) - \rm{sgn}(s') - \rm{sgn}(s'')].
\end{eqnarray}
\end{subequations}
The Hamiltonian is given by
\be\label{eq:hamsecondorder}
\begin{aligned}
{\check H}^{[\le1]}(Q)&=H^{[0]}(Q)+\frac{1}{2} \epsilon \int \! \rmd s \, 
\mathcal{G}_2\big[Q,X^{[0]}_s(Q)\big] +\frac{1}{2}
\epsilon \int
\! \rmd s \rmd s' \, \mathcal{G}_3 \big[Q,X^{[0]}_s(Q), X^{[0]}_{s'}(Q) \big].
\end{aligned}
\ee

\end{widetext}

\section{Infrared divergences in a simple model}\label{sec:appendixA}

This appendix considers scalar fields with a cubic potential, which are sufficiently simple that some of our constructions can be evaluated explicitly. First, we describe at least some of the $n$-point functions that appear in this toy model. We then argue that the iterated time-even conservative sector defined in Sec. \ref{sec:noIEven} is infrared-divergent in this theory. 

The scalar field theory we now consider is assumed to be described by the potential\footnote{Note that this potential is not bounded from below. However, it has a local minimum at $\phi=0$, so the field remains stable as long as perturbations are small.} 

\begin{equation}
    V(\phi)=\frac{1}{2} M^2 \phi^2+\frac{1}{6} \alpha \phi^3,
\end{equation}
where $M > 0$ is the field mass and $\alpha$ is a constant that parameterizes the nonlinearity. We focus on perturbations to the source-free solution $\bar{\phi}=0$ in flat spacetime, so the linearized field operator $\mathbb{D}_{\bar{\phi}}$ reduces to the Klein-Gordon operator $\nabla^\mu \nabla_\mu - M^2$. Eq.\ \eqref{eq:retfieldexpansion} then implies that the first and the second order perturbations to the field must satisfy
\begin{subequations}
    \begin{align}
        \big(\nabla^\mu \nabla_\mu -M^2\big) \phi_{(1)}&=\rho,
        \\
        \big(\nabla^\mu \nabla_\mu-M^2\big) \phi_{(2)}&= \frac{1}{2} \alpha \phi_{(1)}^2.
    \end{align}
\end{subequations}

The advanced and the retarded Green's functions associated with the first of these equations are known \cite{introEMRI}. In terms of the inertial coordinates $(t,\bm{x})$, they have the form
\begin{align}
    G^\pm_2(x,x') = \frac{1 }{4\pi}  \left[- \delta (\sigma) + U (\sigma)  \Theta (-\sigma) \right] \Theta \big( \! \pm (t-t')\big) ,
     \label{GreenKG}
\end{align}
where $\Theta$ is the Heaviside step function and $\sigma(x,x') = \frac{1}{2} \big[ -(t-t')^2 + |\bm{x} - \bm{x}|^2 \big]$ is Synge's world function. Physically, the $\delta(\sigma)$ term in $G_2^\pm$ describes how fields propagate along the null cone, where $\sigma=0$. The $\Theta(-\sigma)$ term instead describes the tail. Using $J_1$ to denote a Bessel function of the first kind, that tail is known to be given more explicitly by
\begin{equation}
    U(\sigma) = M  J_1 (M \sqrt{-2 \sigma } )/ \sqrt{-2 \sigma} .
    \label{Vtail}
\end{equation}
Eqs.\ \eqref{eq:CD2pointCa} and \eqref{GreenKG} show that the conservative 2-point function is
\begin{equation}
    G^\rmC_2(x,x') = \frac{1}{8\pi} \left[ - \delta (\sigma ) + U (\sigma) \Theta
     (-\sigma)  \right]
\end{equation}
in this context.

If we attempt to define a conservative sector by applying the projectors $\psym$ and $\piter$ to the 3-point function $\hat{G}_3^+$, the resulting 3-point function is formally given by Eq.\ \eqref{eq:Cmexp}. That result involves an integral with two types of terms. One type involves integrals over $x'''$ of integrands proportional to, e.g.,
\begin{equation}
    G_{2,1}^\rmS (x,x';x''') \hat{G}^\rmC_2 (x'',x''').  
\end{equation}
Due to the support properties of $G_{2,1}^\rmS$ described in Sec. \ref{sec:2ndOrder}, the integration variable can be restricted to the geodesic segment connecting $x$ and $x'$. This is a finite region, so we do not consider it further. 

The other type of term in the 3-point function given by Eq.\ \eqref{eq:Cmexp} involves an integral over $x'''$ of an integrand proportional to 
\begin{equation}
    G_2^{\rmC}(x,x''')G_2^{\rmC}( x', x''' )G_2^{\rmC}( x'', x''' ).
\end{equation}
As explained in Sec. \ref{sec:noIEven}, the support of this term is generically unbounded at fixed $x$, $x'$ and $x''$, and might therefore diverge. In order to argue that it does in fact diverge in this model, we focus only on that portion of the integral associated with the tail portion of $G^\rmC_2$ and suppose for simplicity that $x$, $x'$ and $x''$ all coincide. The potentially-problematic integral in Eq.\ \eqref{eq:Cmexp} is then proportional to
\begin{align}
    I \equiv \int \! \rmd^4 y \,U^3 (\sigma(x,y) ) \Theta (-\sigma(x,y) ),
    \label{Idef}
\end{align}
which can be integrated by adopting spherical hyperbolic coordinates $(T, \chi, \theta, \varphi)$. Beginning with standard spherical coordinates $(t,r,\theta,\varphi)$ centered on $x$, these are defined via
\begin{equation}
    t = \pm T \cosh \chi, \qquad r = T \sinh \chi,
\end{equation}
where $T = \sqrt{- 2 \sigma(x,y)}$ and $\chi$ are both non-negative. The plus sign covers the interior and the boundary of the $\sigma(x,y) \leq 0$ null cone to the future of $x$. The minus sign covers the corresponding past null cone. The coordinate $T$ is equal to the proper time from $x$ to $y$ along the geodesic that connects them, and the volume element is given by
\begin{equation}
    \rmd^4y=T^3\sinh^2\chi  \sin \theta \rmd T \rmd\chi \rmd \theta \rmd \varphi.
\end{equation}
The integral in Eq.\ \eqref{Idef} therefore reduces to
\begin{align}
    I = 8 \pi M^3 \int_0^\infty \! \rmd T \, [ J_1 (M T) ]^3 \int_0^\infty \! \rmd \chi \sinh^2 \chi.
\end{align}
Although the Bessel function here is oscillatory, the $T$ integral does not vanish. However, the $\chi$ integral diverges exponentially. 

We conclude that the integral over three factors of $G_2^\rmC$ appearing in the iterated time-even 3-point function $\hat{G}^{\Citer}_3$ is infrared divergent at least when the arguments of that 3-point function coincide. More generally, we expect qualitatively similar integrals to arise even when the arguments of this 3-point function are distinct, as their differences should not be significant for the distant integration points that control the divergence. The iterated time-even conservative sector is therefore expected to be generically ill-defined at least in nonlinear theories with finite field mass.

\end{document}